\documentclass[letterpaper,twocolumn,10pt]{article}
\usepackage{usenix-2020-09}

\makeatletter
\def\endthebibliography{%
  \def\@noitemerr{\@latex@warning{Empty `thebibliography' environment}}%
  \endlist
}
\makeatother

\PassOptionsToPackage{hyphens}{url}
\usepackage{hyperref}
   \hypersetup{colorlinks=true, breaklinks=true, linkcolor=blue, urlcolor=blue, citecolor=blue}

\usepackage{cite}
\usepackage[pdftex]{graphicx}
\usepackage[cmex10]{amsmath}
\usepackage{algorithmic}
\usepackage{caption}
\usepackage{comment}
\usepackage{url}
\usepackage{times}
\usepackage{xspace}
\usepackage{multirow}
\usepackage{hhline}
\usepackage{xcolor}
\usepackage{float}
\usepackage{dashrule}
\usepackage[labelformat=simple]{subcaption}
   
\let\labelindent\relax \usepackage{enumitem}
\newcommand{\Sr}{\mathcal{S}}

\newcommand{\Cl}{\mathcal{C}}

\newcommand{\mm}{BAAFE}
\newcommand{\V}{\mathcal{FV}}

\newif\ifshowcomments
\showcommentsfalse

\ifshowcomments
\newcommand{\maliheh}[1]{{\color{cyan}[Maliheh: #1]}}
\else
\newcommand{\maliheh}[1]{}
\fi

\ifshowcomments
\newcommand{\sz}[1]{{\color{green}[SZ: #1]}}
\else
\newcommand{\sz}[1]{}
\fi

\newif\ifshownotes
\shownotesfalse

\ifshownotes
\newcommand{\mynote}[1]{{\color{blue}[Note: #1]}}
\else
\newcommand{\mynote}[1]{}
\fi

\microtypecontext{spacing=nonfrench}

\begin{document}

\title{Privacy-Preserving Application-to-Application Authentication \\ Using Dynamic Runtime Behaviors}

\author{
{\rm Mihai Christodorescu}\\
 \and
{\rm Maliheh Shirvanian}\\
Visa Research
 \and
 {\rm Shams Zawoad}\\
} 

\maketitle


\begin{abstract}

Application authentication is typically performed using some form of secret credentials such as cryptographic keys, passwords, or API keys. Since clients are responsible for securely storing and managing the keys, this approach is vulnerable to attacks on clients. Similarly a centrally managed key store is also susceptible to various attacks and if compromised, can leak credentials. To resolve such issues, we propose an application authentication, where we rely on unique and distinguishable application's behavior to lock the key during a setup phase and unlock it for authentication. 

Our system add a fuzzy-extractor layer on top of current credential authentication systems. During a key enrollment process, the application's behavioral data collected from various sensors in the network are used to hide the credential key. The fuzzy extractor releases the key to the server if the application's behavior during the authentication matches the one collected during the enrollment, \textit{with some noise tolerance}.

We designed the system, analyzed its security, and implemented and evaluated it using 10 real-life applications deployed in our network. Our security analysis shows that the system is secure against client compromise, vault compromise, and feature observation. The evaluation shows the scheme can achieve 0\% False Accept Rate with an average False Rejection Rate 14\% and takes about 51 ms to successfully authenticate a client. In light of these promising results, we expect our system to be of practical use, since its deployment requires zero to minimal changes on the server.  

\end{abstract}


\section{Introduction}
\label{sec:intro}

Authentication of client applications to servers is broadly materialized through a form of credentials, such as API keys and access tokens~\cite{msid,oauth}. In such forms of authentication, it is assumed that the client can securely access and/or manage authentication keys, analogous to username and passwords. Therefore, the responsibility of maintaining and securing the keys is shifted to the consumers to be handled through some form of key management technique \cite{azure}, and \textit{client compromise} is largely ignored in the authentication protocol. The credentials, therefore, cannot be considered fully secure~\cite{google-api, apisecurity, 6973849} since an attacker, who has compromised a client, may obtain the credentials and attempt to authenticate at any time, as shown in recent attacks~\cite{apiattack1,apiattack2, apiattack3}.

Secure management of credentials is also a problem in the context of user authentication. To some extent, biometric and behavioral factors have addressed this issue and made user authentication effortless and secure. 
Unlike user behavior, software behavior has only been used for classification purposes 
(e.g.,  anomaly and intrusion detection) and not application authentication and authorization. 
Fundamentally, the question we answer in this paper is the following:
\begin{quote}
    \emph{Can a secure and privacy-preserving application-authentication system be built given only observations and a model of application behavior?}
\end{quote}
Clearly an authentication system can be created by evaluating the behavioral model on the observed application behavior and, based on the result, accepting the authentication request. This is unsatisfactory as it closely couples the application behavior to the authentication system, raising privacy and confidentiality concerns (the authentication server would see all application behavior) and requiring changes to the authentication server (to replace verification of authentication credentials with evaluation of a behavioral model).


In this work, we design and build an application-authentication technique, referred to as Behavioral Application Authentication using Fuzzy Extractor (\mm), that is resistant to \textit{application compromise}. Our approach continuously collects the behavior of an application, evaluates it using a given classification model, and uses a fuzzy extractor to hide an authentication credential in a fuzzy vault locked with the application's behavior. We hypothesize that just like a human biometrics, application behavior is distinctive and non-replicable (even though noisy) and, hence, could be uniquely mapped to an authentication credential. In the biometric domain, fuzzy extractors~\cite{juels2006fuzzy} have been introduced to handle the mapping between noisy biometric features and a secret key which can then be input into cryptographic protocols, such as an authentication protocol, and we propose to adapt them for application authentication.

Just like fingerprints, application behavior carries a noise; however, unlike fingerprints, in addition to the noise, application behavior may significantly change over time. Therefore, in designing our approach, we should consider variations that happen due to the noise and those that are the result of a more stable change in the application behavior. Selecting the right threshold to overcome the noise is part of this work. This threshold allows the applications to accurately authenticate in different circumstances despite the noise, while rejecting malicious and erroneous attempts. 

Another difference between biometric systems and an ``application behavior'' system is the type of features and the ranges of values each feature can take. While each fingerprint minutiae can be easily encoded into real values, application behavioral features are of different types and are in widely varying ranges. For example, one feature could be the number of packets sent, in the range of thousands or millions, and another feature could be the number of unique destinations in the range of one to ten. 
Wide variety of features and ranges of values they acquire, demand an encoding scheme that maps raw feature values to real numbers in a certain range, while maintaining the noise tolerance before and after the encoding. Hence, we introduce normalization techniques and an encoding scheme, whose impact on accuracy and security we empirically measure in our evaluation.

\begin{figure}
	\centering

	\begin{subfigure}{\columnwidth}
		\raggedright
		\includegraphics[width=0.88\textwidth]{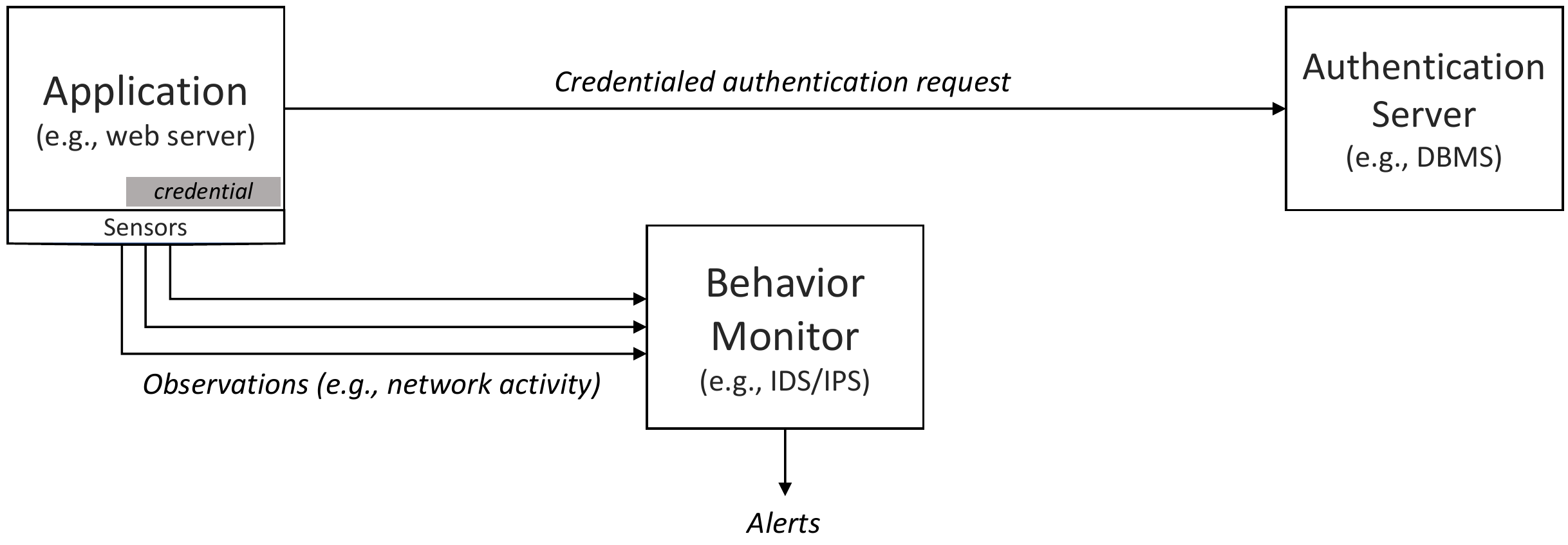}
		\caption{A traditional authentication architecture places the credential on the side of the application client.}
		\label{figure:traditional-auth}
	\end{subfigure}

	\vspace{\baselineskip}

	\begin{subfigure}{\columnwidth}
		\includegraphics[width=\textwidth]{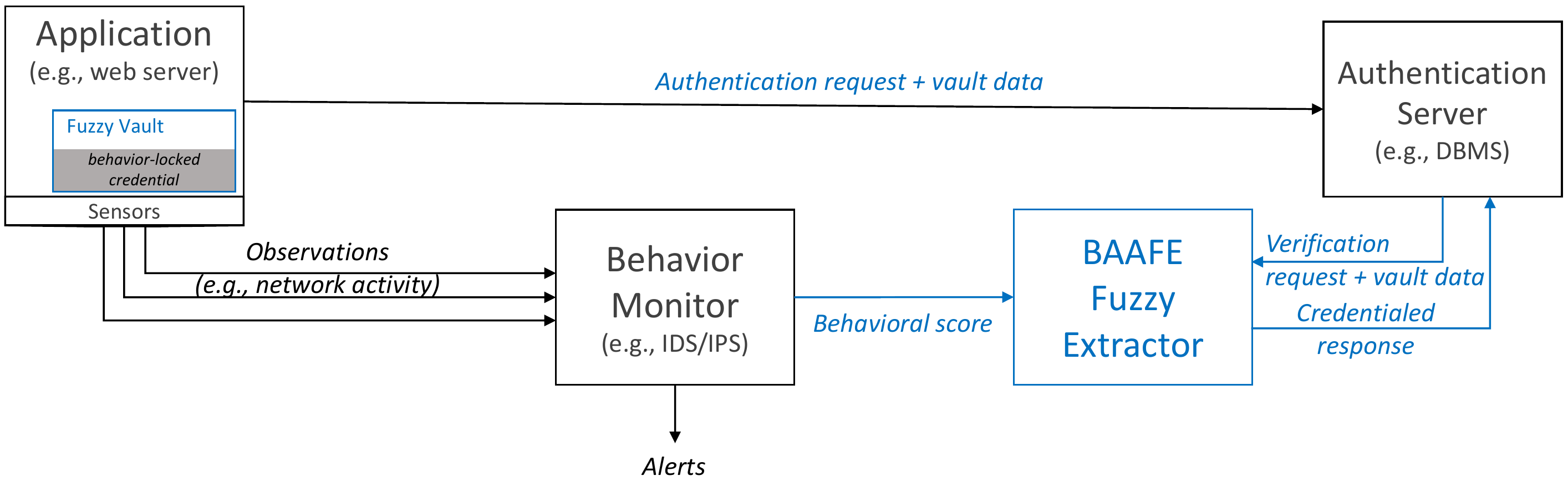}
		\caption{The BAAFE architecture locks the client-side credential in a ``fuzzy vault,'' unlocked only by the correct application behavior.}
		\label{figure:baafe-auth}
	\end{subfigure}

	\caption{Comparison of traditional and BAAFE authentication architectures.}
	\label{fig:overview}
\end{figure}

\autoref{fig:overview} shows a conceptual view of the proposed system. 
In current authentication models (\ref{figure:traditional-auth}), application behavioral data collected by sensors, such as software and operating system event logs and network firewalls, are evaluated by a behavioral model to detect honest application behavior. We introduce a fuzzy-extractor into the current design (\ref{figure:baafe-auth}). 
During an enrollment process, \mm\ secret-shares the authentication credential key as coefficients of a polynomial and hides the key on a public vault in a privacy-preserving form by adding random chaff points. 
Application behavior are encoded into numerical values and lie on the polynomial, while chaff points are randomly distributed outside the polynomial.  
During authentication, \mm\ unlocks the fuzzy extractor vault and releases the key to the server if it can correctly reconstruct the polynomial from the current application behavior data. \mm\ tolerates a certain level of noise, just enough to authenticate different attempts by the same application while rejecting other malicious and non-malicious applications.


We do not build a server-side behavioral recognition system, rather we show how to layer the \mm\ fuzzy extractor on top of an existing application behavior classifier such as an IDS/IPS. This has the main benefit of requiring zero to minimal changes on the server while resisting client-side attacks, as the server will still authenticate the applications using commonly deployed credential authentication protocols.
Additionally, the server does not learn anything about the behavior of the client, which makes it a privacy-preserving approach. This is a desirable characteristic for use-cases where an external organization such as cloud platforms offer services through their API's (e.g., Amazon Web Service \cite{aws} or 
Google Cloud Platform \cite{google-cloud}).

\vspace{0.25\baselineskip} \noindent \textbf{Contributions.} The contributions of this work are as follows: 

\begin{itemize}[itemsep=0.25\baselineskip, leftmargin=*]

\item{Design of a privacy-preserving authentication based on application behavior.}
To realize the idea of incorporating dynamically changing application behavior in the  authentication process, we present the design of a system for Behavioral Application Authentication using Fuzzy Extractors (BAAFE) that is secure against client compromise, vault compromise, and feature eavesdropping.


\item{Novel characterization of application behaviors for use with fuzzy extractors.} Although application behaviors and biometrics have many similarities, fundamental distinctions mean that existing approaches to fuzzy extractors on biometrics fail for application behaviors. We describe techniques to bridge this gap, showing how to adapt fuzzy extractors for application behaviors and demonstrate its feasibility.

\item{Implementation and evaluation of BAAFE accuracy and efficiency.} We implemented our authentication scheme for 10 real-world applications running in a large network environment. Our evaluation over one month of data suggests a promising result with an accuracy of 0\% False Reject Rate (FRR) for five applications, an average FRR of 14\%, and an authentication latency of 51 milliseconds.

\end{itemize}

\section{Background and Prior Work}
\label{sec:background}

\vspace{0.25\baselineskip} \noindent \textbf{Behavioral Biometric Authentication.}
In the context of user authentication, behavioral authentication refers to techniques that can authenticate a user solely based on some quantifiable and unique users' behaviors. Support of various types of sensors on smartphones and wearable devices have encouraged this type of authentication due to its implicit and easy user interaction model.  The idea is to find features of behavior that can uniquely identify a user. A model can then be trained on these features to authenticate the user for each session (or keep the user authenticated in case of continuous authentication). Some examples of behavioral authentication researched in the past include keystroke dynamics, mouse movement, and walking patterns. As an example, HMOG \cite{sitova2015hmog} uses Hand Movement, Orientation, and Grasp data collected  for continuous user authentication on smart phones. In \cite{sitova2015hmog}, features are collected from touchscreen, accelerometer, gyroscope, and magnetometer and Scaled Manhattan (SM), Scaled Euclidean (SE), and Support Vector Machine (SVM) are used for classification. SilentSense \cite{bo2013silentsense}, suggests authenticating smartphone users from the user touch behavior biometrics and movements of the device caused by touchscreen actions. iAuth \cite{lee2016implicit} uses the same sensors, accelerometer, and gyroscope on both smartphone and smart watch for authentication. 

\vspace{0.25\baselineskip} \noindent \textbf{Machine Learning for App Classification and Device Fingerprinting.} 
Hardware and software classification according to device and application behavior is not a new topic in the area of deep learning and machine learning. 
Behavioral features have been suggested in software and application traffic classification. In this rich line of research, software behavior collected by network devices such as intrusion detection and firewalls can detect and block malicious activities. In  \cite{lane1997application}, the system learns normal user behavior and classifies current behavior as normal or anomalous according to their similarity to past behavior. In a book written by Bhattacharyya and Kalita \cite{bhattacharyya2013network}, several features and classification techniques are described for network anomaly detection using machine learning. Some of the features collected in our experiment Section \ref{sec:study} are inspired by this selection. While behavioral features have been used in the past to successfully classify applications, they have not been used as an authentication factor. Many of the proposed techniques perform a binary classification to identify malicious software, network traffic, or transactions according to the behavior of the application, e.g., a survey on  deep learning for anomaly detection \cite{chalapathy2019deep}. Others are multi-class solutions to classify type of traffic or application, e.g., \cite{zander2005automated, jiang2007lightweight, lin2009application}. 

Another related  is device fingerprinting in which the behavior of a hardware device, e.g., a client machine, a wireless access point, or an IoT device, is used to identify or authenticate the device. Xu et al. \cite{xu2015device}, for instance, targets wireless device fingerprinting. They recognize and introduce several features in development on a machine learning algorithm to identify a wireless device.  IoTSense \cite{bezawada2018iotsense} introduced a method to identify IoT devices from a small number of network packets. They noticed that certain device network activities such as TCP window size, entropy,  and payload lengths, are specific to a device and can accurately identify a device. We can use this technique in identifying a hosting machine. 

While prior work has shown immense success in the classification context, their application in the client-to-server authentication domain has not been explored in the past. Our work tried to fill the gap between classification of an application and authentication of an application to service using traditional credential-based authentication. We identify unique and distinguishable hardware, software, and network features and show a fuzzy extractor construction that can map these features to a secure authentication credential. 

\vspace{0.25\baselineskip} \noindent \textbf{Fuzzy Extractor.}
Fuzzy extractor is a cryptographic construction that allows biometric embedding to be used in cryptographic protocols. Since cryptographic techniques work with fixed and deterministic values fuzzy extractors suggest techniques to tolerate the noise in the biometric data. User authentication is one of  the most natural applications of fuzzy extractors working on fingerprint, face, and iris \cite{juels2006fuzzy, wang2007fuzzy, sutcu2009design, tong2007biometric, marino2012crypto, dodis2008fuzzy}. In this application, the user's biometric data is used to decommit a cryptographic key that could be used just like any other cryptographic key in the authentication process. From the usability perspective, the advantage of using fuzzy extractors is that biometric data, which is more inherent and hence usable authentication factor, could be adopted by any secure authentication scheme. From the security and privacy point of view, biometric templates are not stored on the servers and the key could frequently be updated, unlike biometric data.

\emph{Informal Definition:}  Fuzzy extractor is a pair of a generation ${Gen}$ and reconstruction ${Rec}$ procedures, where for a biometric fuzzy input $b$, ${Gen}$ extracts a t-bit private string  $s \in \{0,1\}^t$ and a public helper data $h \in \{0,1\}^l$ where $l>>t$. ${Rec}$ on the input of a biometric fuzzy input $b'$ and $h$, outputs $s$ if the distance between $b$ and $b'$ is less than a given threshold $\tau$, i.e., $d(b,b')<\tau$.  

\emph{Basic Fuzzy Vault Construction:}  The first construction of Fuzzy Extractor called Fuzzy Vault \cite{juels2006fuzzy} suggests that to lock a vault a set of codewords representing $s$ as a polynomial $P$ of degree $k$ (e.g., $P$ has an embedding of $s$ in its coefficients) should be created. Assuming that $b$ can be encoded as a set of features $F$, each codeword is shown as a  point $(x_i, P(x_i))$ where x coordinates $\in F$ and y coordinates are the projection of $x_i$ on $P$. To conceal the point random chaff points $(c_i \neq x_i, y_i \notin P(c_i))$ are added to codewords. Actual codewords and chaff points are published as the helper data or vault. 
To unlock the vault, upon receiving $b'$, sets of points $\in h$ whose x coordinates are close enough (according to the distance function) to the input features are selected and P is reconstructed. $k+1$ provided features should be close enough to the features stored in the vault for the polynomial to be reconstructed and $s$ being revealed.



\section{Overview}
\label{sec:model}


We consider a client-server environment in which a client application (a.k.a. client\footnote{For brevity throughout this paper we use the term \textit{client} to refer to a client application, which operates without any involvement from a human. In particular, there is no human available to provide biometric data or other kinds of interactive authentication data.}) that resides on a client machine requests and receives services from a server application. The client needs to 
be authenticated for the request to be served. An example of such interaction is 
a web-application server sending a query to a database server. 
A common authentication approach is to use some form of credentials such as API keys, 
tokens, or passwords, stored on the client and verified on the server.
However, managing credentials in a secure manner is challenging, particularly on the 
client side.

We aim to make the client-to-server authentication secure and more inherent by incorporating \textit{behavior} of the client into the authentication process. Client application behavior is used as an application ``biometric'' in a fuzzy extractor construction to hide a given cryptographic key in a vault and retrieve it during authentication.
Our proposed 
behavioral fuzzy extractor authentication scheme aims to bind the credential 
(from now on referred to as key to cover tokens, password, a security/cryptographic keys)  to the 
behavior of the client. 
While this approach incorporates the behavior of the client in the 
authentication process, it does not attempt to deploy a behavioral model at the server, as is the case with intrusion and anomaly detectors. Rather, server-side authentication 
is unchanged and can use well-established and secure protocols that rely on 
credentials. 
Therefore, key management would be easier on the client, does not require trust on the client, and 
is as secure as credential authentication on the server. 

The parties involved in the protocol are a client application $\Cl$ known by an application id \texttt{app$_{id}$} trying to authenticate 
to a server application $\Sr$. The client $\Cl$ only provides its behavior $B$, and the server only accepts a secret key $k$ as authentication proof. 
The secret key $k$ is hidden in a vault $V$ stored publicly, for example, on $\Cl$. A Fuzzy Extractor Service $\V$ can reconstruct the secret from the vault using application behavior $B$.
Practically, 
the server $\Sr$ itself can host the fuzzy extractor service $\V$, however, we treat them as 
separate components for three reasons: 1) the server and fuzzy extractor could belong 
to different organizations, 2) one fuzzy extractor could serve multiple servers, and 3) the design and security model are more clear.

Our system, called \textbf{\mm} for \emph{Behavioral Application Authentication using Fuzzy Extractors}, consists of two protocols,  
an \textit{enrollment} protocol, $V= Enr(B,k)$, to bind a 
cryptographic key $k$ to the behavior $B$ of the client and store it in a vault $V$, and 
an \textit{authentication} protocol, $k=Rec(B,V)$ to recover the key $k$
from the vault $V$ on observing client's behavior $B$. The key $k$ can then be used to authenticate to the server $\Sr$. 

We refer to the following terms in the rest of this paper.

\vspace{0.5\baselineskip}	
	\noindent\textbf{Feature:} A single property related to a client's execution or its client-to-server interaction. Each feature can get a value at any time that can be encoded into a numerical representation using an encoding technique. While a set of features are unique and non-replicable over time, an individual feature might be the same/similar among multiple applications or instances of the same application. Some examples of features include host-based features (e.g., average daily application CPU consumption), application audit trail-based features (e.g., number of successful connections per day), and network-based features (e.g., number of unique URLs per day).

\vspace{0.5\baselineskip}	
	\noindent\textbf{Sensor:} A mechanism to collect the instantaneous value of features associated with an application. Sensors can include firewalls, intrusion detection systems, and operating-system monitors. Sensors provide features values at any given time and may store or track the history of these values. 
	
\vspace{0.5\baselineskip}	
 	\noindent\textbf{Behavior:} A sequence of feature values gathered over time from one execution of the application. 
 	
\vspace{0.5\baselineskip}	
 	\noindent\textbf{Behavioral Model:} The model maps the behavior to a risk score. The features that make up the behavior must be sufficiently informative to allow the operation of a machine-learning classifier that distinguishes normal from compromised executions in an accurate and robust manner. In this context robustness means resilience to adversarial attacks and to mimicry attacks~\cite{10.1145/586110.586145} in particular. This places a lower bound on the number and types of features collected from an application execution.


\subsection{Design Goals}

We considered four goals while designing the system.

\begin{enumerate}[leftmargin=*,noitemsep]

\item{\textbf{Security against  Client Compromise.}} An attacker who compromises a client application
cannot authenticate to the server unless it allows the client application to continue to execute normally.

\item{\textbf{Security against Vault Compromise.}} An attacker who gets access to the fuzzy vault data cannot 
learn the key used to authenticate to the server. 

\item{\textbf{Security against Feature Observation.}} An attacker who observes the execution of the client application (via features on the client host or on the network) cannot learn the key used to authenticate independently to the server.

\item{\textbf{Compatibility with Server Credentials.}} The system should not require changes to the server-side
authentication. Servers can still use credentials and credential-based protocols to authenticate clients. 
This is particularly important when the client application and the authentication server are part of different domains, as is the case with application designs which rely on SaaS applications.

\end{enumerate}


\subsection{Threat Model}

We consider an attacker who can compromise the client application and its runtime environment. This means the attacker can change the execution of the application, inject code and data, and block incoming and outgoing requests.
The attacker can also read but not modify the fuzzy vault storage and can passively read the sensor measurements. The attacker can be adaptive (e.g., to perform mimicry attacks) and can use data exfiltrated in an earlier compromise.

\vspace{0.25\baselineskip}\noindent\textbf{Trusted components and setup.}
We consider the server, the fuzzy extractor (including the encoding parameters), the sensors/behavioral model, and the communication channels between them to be secure. The server and the fuzzy extractor are trusted to be safe from compromise by an attacker. Sensors are trusted to report accurately on the client application execution, regardless of whether that execution is normal or anomalous.


We assume that during the enrollment/setup phase all parties, including the client application, are trusted and their communication is secured. That is, 1) the client is not malicious nor compromised and sends a valid request to the server, 2) server runs a secure key exchange protocol to transfer the client's key to fuzzy extractor, 3) the attacker is not present to observe sensor data and the data collected during the enrollment is available only to the fuzzy extractor. Additionally all cryptography primitives used in our construction are assumed to be secure.

\vspace{0.25\baselineskip}\noindent\textbf{Example settings.}
The simplest instantiation of this model is a client application running on a host and a set of sensors in a network firewall observing the network communication between that host and a cloud service. In this setting a compromise gives the attacker full control of the application, of its code and data in memory, and of the host software and hardware itself, but not of the sensors nor the server.
Other instantiations include placing the sensors in the OS kernel to collect system-call events (and would protect only against user-space compromise) or in a hypervisor (thus protecting against both user- and kernel-space compromise).

\section{Enrollment and Authentication Protocols}
\label{sec:system}
In this section, we present the key enrollment process and the algorithm as well as the  key reconstruction and authentication process flow and the related algorithm. Let us consider a (public) finite field over which all algebraic operations are carried out, and the following security parameters:
\begin{description}[topsep=0pt, noitemsep] 
\item[\textbf{$n$}:] number of features selected from the application behavior to be included in the vault;
\item[\textbf{$c$}:] size of the chaff vector in the vault;
\item[\textbf{$d$:}] degree of the polynomial whose coefficients are
shares of the authentication key $k$.
\end{description}

The protocol relies on several building blocks. 

\begin{itemize}[itemsep=0.25\baselineskip, topsep=0.25\baselineskip, leftmargin=*]
\item \textbf{Hash Function.} A hash function $H$ is a one way function that maps an arbitrary size string to a fixed-size value (e.g., SHA256). 
\item \textbf{Secret Sharing.} A secret sharing algorithm $SS$ that breaks a secret $k$ into $d+1$ shares. 
Each share  carries only part of the secret and does not reveal the secret on its own. 
For our purposes, all of the shares must be combined to reconstruct the secret.
\item \textbf{Normalization.} The ranges of values the features acquire varies from one feature to another, so normalization projects these values into the same range, to enable comparisons.

\item \textbf{Encoding.} 
We define an encoding function $E$ as a deterministic two way function that maps feature values to values fitting fuzzy vault distance function and threshold requirements.  

\item \textbf{Distance Function.} The distance between two feature values $x$ and $x'$
can be measured by a distance function $D(x,x') = |x-x'|$. We used euclidean distance to measure the distance between two points. The closeness of a feature value collected during the authentication to the one stored in the vault is computed using the distance function.

\item \textbf{Threshold.}  If the distance is less than a predefined threshold, the vault value is selected to reconstruct the secret. Overall, a larger threshold allows higher noise therefore reduces the false rejection rate, but may increase the chance of authenticating wrong applications, i.e., increases the false acceptance rate. Section \ref{sec:challenges} describes various approaches of measuring threshold and the experimental results presented in Section \ref{sec:study} shows the effect of thresholds on the accuracy and security.

\end{itemize}


\subsection{Enrollment}


\begin{figure}

	\begin{subfigure}{\columnwidth}
		\centering
		\includegraphics[width=0.96\columnwidth]{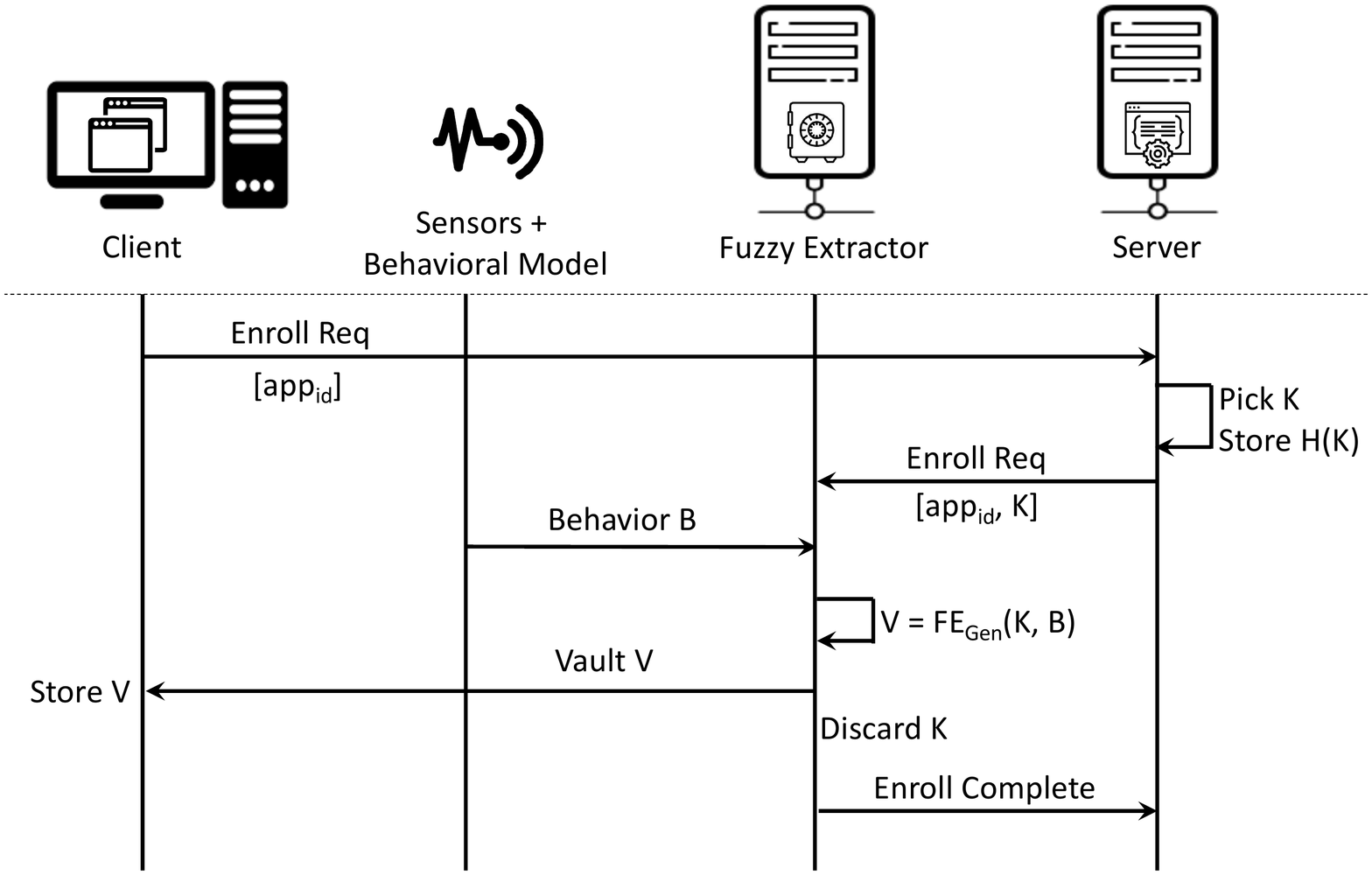}
		\vspace{-8mm}
		\caption{Enrollment Workflow}
		\label{figure:enrollment_flow}
    \end{subfigure}
    \vspace{2mm}

	\begin{subfigure}{\columnwidth}
		\begin{tabular}{@{}p{0.96\columnwidth}@{}} 

			\rule{0.96\columnwidth}{0.5pt}

			\noindent Input: behavior $B$ from $\Cl$, key $k$ from $\Sr$

			\noindent Output: vault $V$

			\hdashrule[0.5ex]{8.2cm}{1pt}{1pt}

			\begin{itemize}[noitemsep,leftmargin=*]

				\item $\Cl$ sends an enrollment request to $\Sr$ as \texttt{app$_{id}$}

				\item $\Sr$ generates and stores a key $k$ and sends ($k$, \texttt{app$_{id}$}) to $\V$

				\item $\V$ proceeds as follows:

					\begin{enumerate}[noitemsep]
						\item Let $P_{d}(x)=c_dx^d+...+c_1x+c_0$, where $\{c_i\}$ is a secret-sharing of $k$
						\item Encode $B$ to $X = \{x_1,...,x_n\}$
						\item Compute $n$ genuine points: $$F = \{(x_i,P_d(x_i)) : 1\leq i\leq n \}$$
						\item Generate $c$ random chaff points: $$C = \{(r_{x_i}, r_{y_i}) : 1\leq i\leq c, r_{x_i} \notin X, r_{y_i} \neq P_d(r_{x_i})\}$$ 
						\item Vault is initialized as $V = \langle \mathtt{app}_{id}, C \cup F, H(k) \rangle$ 
					\end{enumerate}

			\end{itemize}

			\rule{0.96\columnwidth}{0.5pt}
		\end{tabular}
		\vspace{-2mm}
		\caption{Enrollment Protocol}
		\label{fig:enrol}
	\end{subfigure}

    \vspace{2mm}
	\begin{subfigure}{\columnwidth}
		\centering
		\includegraphics[width=.96\columnwidth,trim=0 45 0 0, clip]{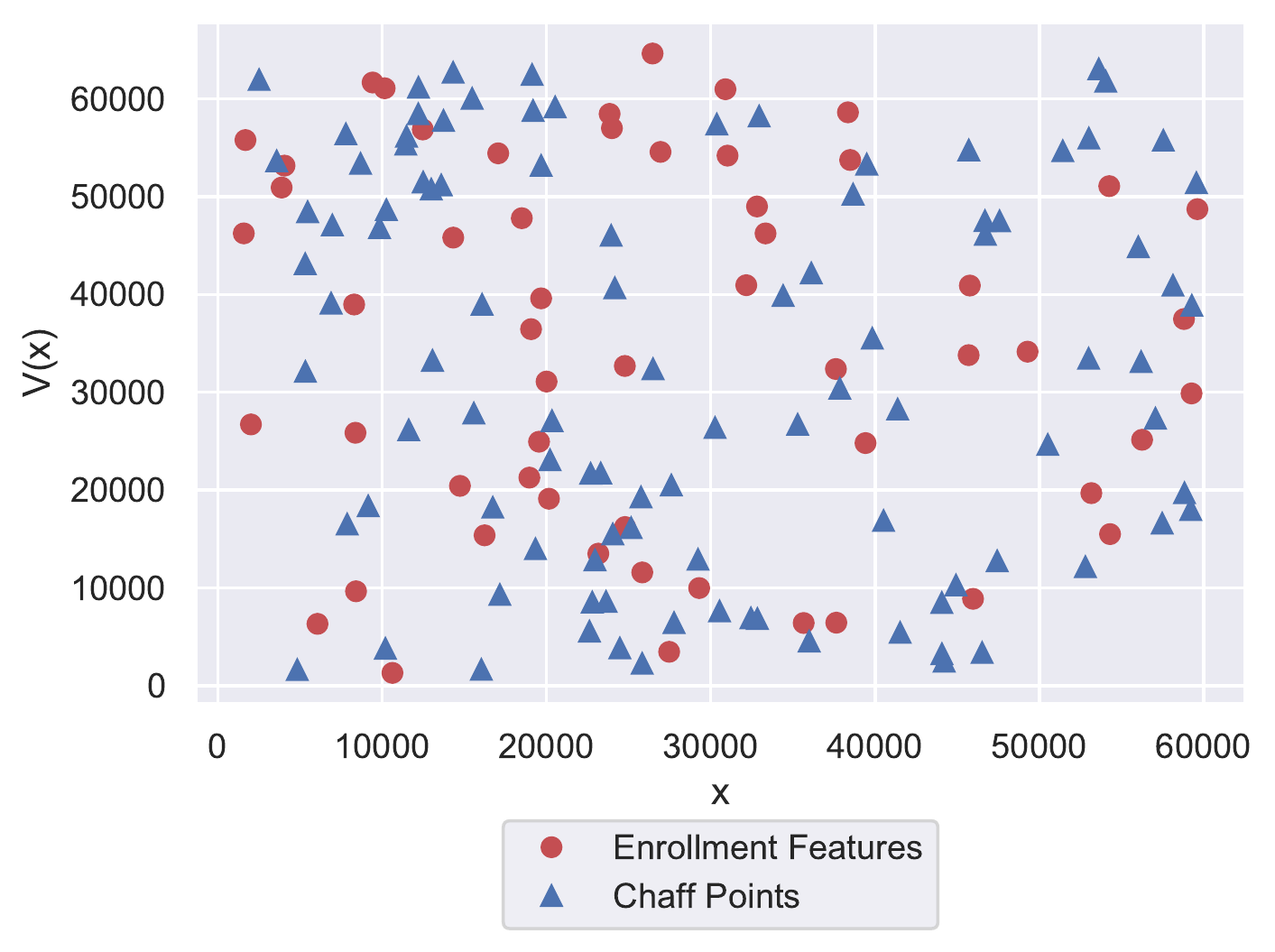}
		\caption{Points in fuzzy vault. {\color{red}\Large\textbullet} = genuine, {\color{blue}\Large\textbullet} = chaff}
		\label{figure:vault_viz}
	\end{subfigure}

	\caption{\mm\ enrollment}

\end{figure}

\autoref{figure:enrollment_flow} illustrates the key enrollment process and various interactions between the components.  To start the enrollment process, the client sends the key enrollment request  to the server with a new application ID. The server chooses the key, stores the Hash of the key and forwards the request to the Fuzzy Extractor (FE) with the key. The Fuzzy Extractor then extracts the application's behavioral data from the behavioral model and prepare the fuzzy vault using the enrollment algorithm described in \autoref{fig:enrol}. The enrollment algorithm takes the behavioral data and the key as input and produce the fuzzy vault as output. The fuzzy extractor send the fuzzy vault to the client to be stored, removes the key and sends a completion acknowledgment to the server. \autoref{figure:vault_viz} shows a visual representation of the fuzzy vault with enrollment features and chaff points. One of the key step of enrollment process is step 4, where the system adds chaff points to vault. When adding chaff points, the algorithm makes sure that the chaff points are outside of the threshold region from all the original points and are not on the polynomial.



\subsection{Key Reconstruction and Authentication}

\begin{figure}

	\begin{subfigure}{\columnwidth}
		\centering
		\includegraphics[width=0.96\columnwidth]{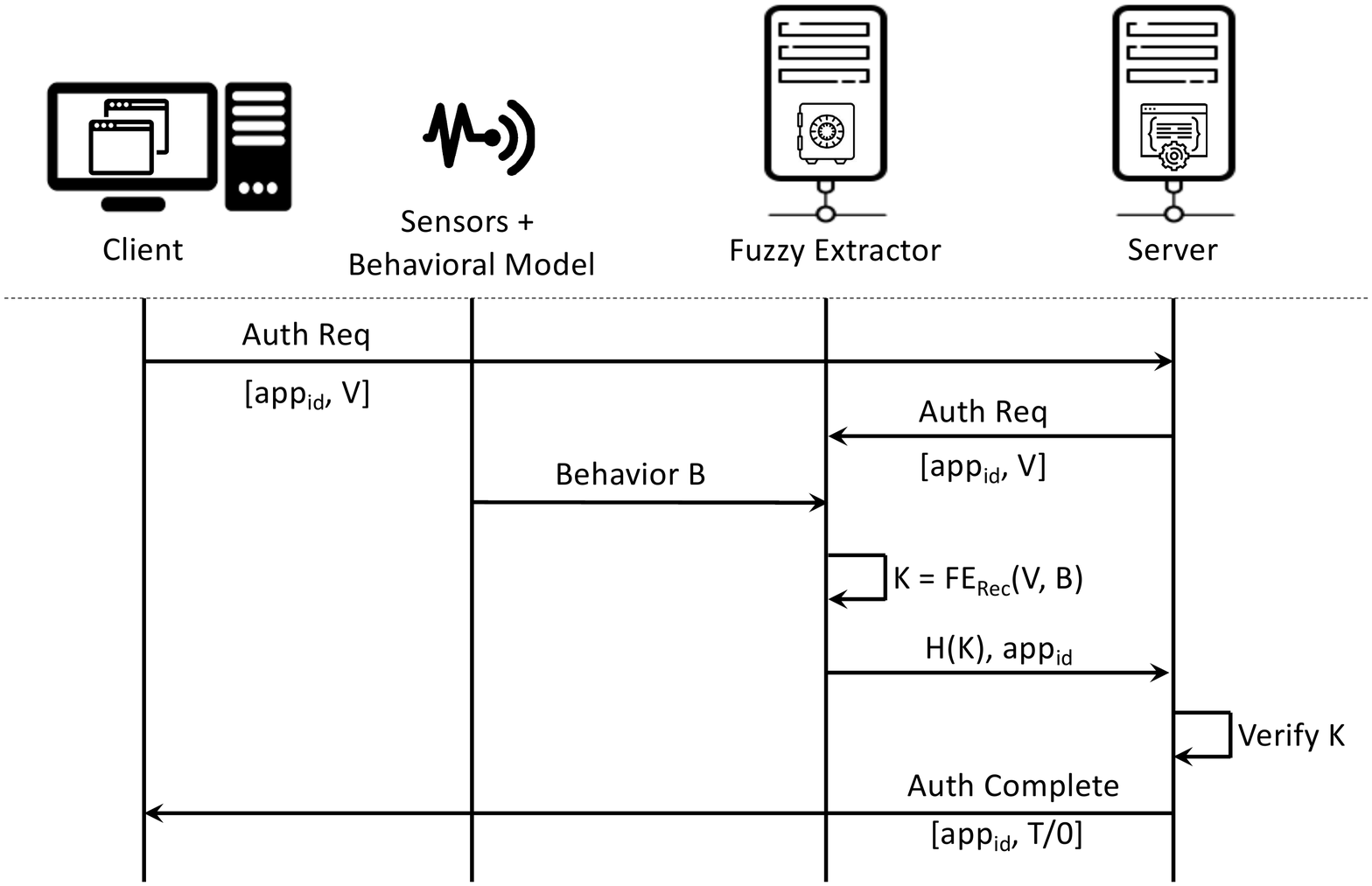}
		\caption{Authentication Workflow}
		\label{figure:recon_flow}
	\end{subfigure}

	\begin{subfigure}{\columnwidth}
		\begin{tabular}{p{0.96\columnwidth}}

			\rule{0.96\columnwidth}{0.5pt}

			\noindent Input: behavior $B'$, vault $V$, threshold $\tau$

			\noindent Output: on success, key $k$; on failure, $\bot$

			\hdashrule[0.5ex]{8.2cm}{1pt}{1pt}

			\begin{itemize}[noitemsep,leftmargin=*]
				\item $\Cl$ sends an authentication request (\texttt{app$_{id}$}, $V$) to $\Sr$. 
				\item $\Sr$ forward the request to $\V$ as a challenge.  
				\item $\V$ proceeds as follows to respond to $\Sr$:  

				\begin{enumerate}[noitemsep]

					\item Using the encoding function, behavior $B'$ is encoded into $X' = \{x'_1,...,x'_n\}$.

					\item  For each element in the encoded behavior vector, $x'_i \in X', 1 \leq  i \leq n$, the closest pair $(v_{x_i},v_{y_i}) \in V$ is selected if $D(v_{x_i}, x'_i) \leq \tau$. 

					\item A candidate set $F' = \{(v_{x_1},v_{y_1}), ..., (v_{x_q},v_{y_q})\}    , d+1 \leq  q \leq n$ is generated for all the pairs selected from the vault representing possible points on the polynomial. Protocol outputs $\bot$ if size of  $F'< d+1$. 

					\item For a number of possible attempts $<$ $ q \choose d+1$ client proceeds with the following and aborts with $\bot$ if maximum number of attempts is reached: 
					\begin{itemize}
						\item  $(d+1)$ elements of $F'$ are selected to interpolate a polynomial $P'_d(x)=c'_dx^d+...+c'_1x+c'_0$.
						\item Key $k'$, is reconstructed using the $(d+1)$ shares $c' = \{c'_0,...,c'_{d}\}$, driven from the interpolated polynomial $P'_{d}(x)$. 
						\item If $H(k') = H(k)$ respond with $k$, else repeat step 1-3 until the maximum number of attempts is reached. 
					\end{itemize}

				\end{enumerate}

				\item $\Sr$ authenticates $\Cl$ if the respond received from $\V$ matches $H(k)$.   

			\end{itemize}

			\rule{0.96\columnwidth}{0.5pt}

		\end{tabular}
		\caption{Authentication Protocol}
		\label{fig:recon}
	\end{subfigure}

	\caption{\mm\ authentication}

\end{figure}

Figure \ref{figure:recon_flow} illustrates the work flow for authentication process for the proposed \mm\ system. 
To authenticate to a server, the client sends an authentication request with a application ID and the stored vault to the server. The server forwards the authentication request to the fuzzy extractor, which extracts the behavioral features of the application from the behavioral model. The fuzzy extractor then runs the key reconstruction algorithm described in Figure \ref{fig:recon} to reconstruct the key from the current behavior of the application and the fuzzy vault. At this point, the fuzzy extractor sends the hash of the extracted key to the server, where the server verifies the key. If the hash of the reconstructed key and the stored key are matched, the server sends an authentication success request to the client.  Figures~\ref{figure:vault_success_viz} and~\ref{figure:vault_failed_viz} in the Appendix give a visual representation of the close point matching during key reconstruction for a successful and unsuccessful attempt.

\section{An Empirical Evaluation of Feasibility}
\label{sec:study}

While the protocols of \autoref{sec:system} sketch a potential authentication system, applicability to real-world settings hinges on whether application behavior is sufficiently ``biometric like'' to allow the use of fuzzy extractors.
Here, we detail our experiments with several applications and their behaviors and discuss the challenges observed and the lessons learned in trying to identify optimal behavioral-authentication thresholds.

\subsection{System Configurations}

\noindent\textbf{Applications.}
To evaluate our proposed solution, we picked 10 applications running in the enterprise systems. The nature of the applications ranges from antivirus, automation engine, malware sandboxing to endpoint response tools. Among these 10 applications, there are two applications, which are running on two different environments but providing the same service. We considered them as two separate applications because the same application's behavior might be different on two different environments and the keys for authentication can also be different for different operational environments. 

\noindent\textbf{Features.}
 We collected 14 behavioral attributes for each of the application from various sensors of the enterprise network, such as enterprise data lake, Security Incident and Event management System (SIEM) which store logs from all NIDS (Network intrusion detection system), firewall logs, network flow data, internet proxy logs, and more. After applying statistical measurements on the data points, we then calculated 56 features for each application. The behavioral attributes are listed in the appendix (Section \ref{sec:appendix})

We collected two months of data for each of the above mentioned attributes and calculated the mean, standard deviation, 50th percentile (median), and 75th percentile value for each attribute for 15 days time window, which resulted into total 56 features for each application.

\subsection{Optimal Threshold and Threshold Scheme Selection}
We consider three approaches for selecting a threshold.
\begin{enumerate}[itemsep=0.25\baselineskip,leftmargin=*]

    \item \textbf{One global threshold}:

    In the first approach, we used one single threshold for all features and all applications. For this approach, we started with a numeric value and found the optimal value by gradually changing the threshold in our experiments.

    \item \textbf{One threshold per application instance}:

    In the second approach, one single threshold is selected for all features of one single application. For this approach, we first normalized the two months of behavioral data of an application using MinMax Scaler algorithm. Then we encoded the normalized values using the encoding method discussed earlier and calculated the $b_1$ and $b_2$ components for each feature. For each feature, we then identified the minimum and maximum values of $b_1$ and $b_2$ components and calculated the difference between the maximum and minimum values. Finally, we selected the maximum difference from all the features of an application to use that as the threshold for the application. We used different configuration for the difference between the maximum and minimum values of $b_1$ and $b_2$, such as $\frac{\max-\min}{2}$ and $\frac{\max-\min}{4}$, and identified the optimal configuration based on our experiments.

    \item \textbf{One threshold per feature of each application instance}:

    In the third approach, we labeled each feature and select a threshold vector consisting of specific thresholds for each feature based on the ranges of values that feature will take. Similar to the second approach, after data normalization and encoding, we calculated the minimum and maximum values of $b_1$ and $b_2$ components for each feature. We used the difference between the maximum and minimum values of $b_1$ and $b_2$  and various configuration such as, \begin{math}\frac{\max-\min}{2}\end{math}, \begin{math}\frac{\max-\min}{3}\end{math}  as the threshold of the feature and found the optimal configuration based on experiments.
    
    Having one threshold per feature of each application  changes the enrollment and authentication protocols as follows.

    Step~4 of the enrollment process, where the chaff points are generated, the system produces a set of chaff points for each feature for this threshold scheme, resulting in $n + n \times c$ total number of points in the vault, where $n$ is the number of features and $c$ is the number of chaff points for each feature. Besides that, while adding the chaff points for a feature, the system makes sure that the chaff points are outside of the threshold region of that particular original feature point. This process is different than the other two threshold schemes, where we check the threshold boundary with all the original feature points while adding a chaff point.
    
    Similarly, step~2 of the authentication protocol (finding the closest pairs) works differently for per feature of each application threshold scheme. For this threshold scheme, for each element of the encoded behavior vector, rather than checking the threshold with all the points of the vault, the algorithm finds the closest pair by comparing the distance only with the points that are labeled for that specific behavioral feature.

\end{enumerate}

In this section, we explain how to choose the optimal threshold in each of the approaches and then empirically identify the best performing approach for further performance evaluation.

\subsubsection{Selection Criteria}
To use the proposed model for application authentication, it is important that the system has low FAR (false acceptance rate) and low FRR (false rejection rate). Hence, for the optimal threshold selection and choosing the best threshold technique, we focused on the FAR and FRR. Our target is to get the lowest FAR and FRR possible for any configuration.

\textbf{FRR.} To measure the FRR for a certain system configuration, we first generated the authentication data for each application by sliding the time window by one day and measures the statistical data for each of the features for a 15 days window. In this way, we generated 15 authentication dataset  for each of the applications. We use each of these dataset to authenticate using the respected application's vault.  If the authentication failed for a dataset, we counted that as a false rejection. Then FRR is calculated as the percentage of total number of authentication attempts and  false rejection. 

\textbf{FAR.} To measure the FAR for a system configuration, first, we picked a random application (Application 1) and completed the enrollment process for the application. Then we used the other 9 application data to authenticate as Application 1. If the authentication succeeded, we counted that as a false acceptance. Then FAR is calculated as the percentage of total number of authentication attempts and false acceptance.

\subsubsection{Optimal Threshold Selection} 
\begin{figure}
\centering
\includegraphics[width=.7\columnwidth]{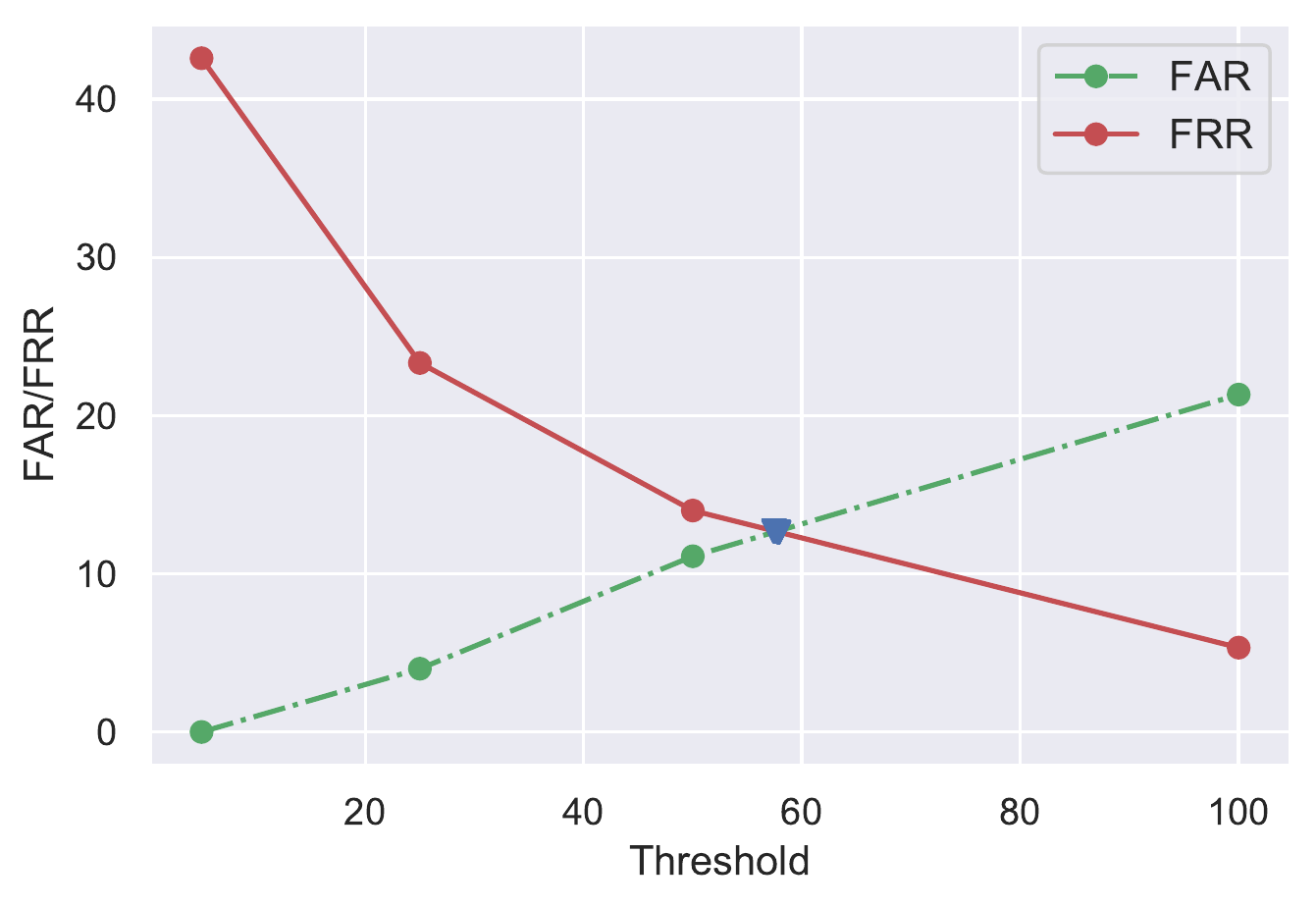}
\caption{Optimal threshold selection for single-threshold technique.}
\label{figure:optimal_threshold}
\vspace{-5pt}
\end{figure}

For one global threshold for all applications approach, we used thresholds 5, 25, 50, 100 with 32 degree polynomial configuration and measured FAR and FRR of the system. The results are shown in \autoref{figure:optimal_threshold}. The point where FRR and FAR line crossed, we chose that as the optimal threshold, which was 57.5 in this case.

For the one  threshold  per  application  instance approach, we studied the application data for two months and calculated threshold with various formulas related to maximum and minimum values of $b_1$ and $b_2$ components of each feature, such as \begin{math}{Max-Min}\end{math}, \begin{math} \frac{Max-Min}{2}\end{math}, \begin{math} \frac{Max-Min}{3}\end{math}, and \begin{math} \frac{Max-Min}{4}\end{math}. We then measured the FAR and FRR for each of the configurations and found \begin{math} \frac{Max-Min}{3}\end{math} generated lowest FRR and FAR.

Similar to the application specific threshold approach, we used different formulas to calculate the threshold for one threshold per feature of each application instance approach: \begin{math}{Max-Min}\end{math}, \begin{math} \frac{Max-Min}{2}\end{math}, \begin{math} \frac{Max-Min}{3}\end{math}, and \begin{math} \frac{Max-Min}{4}\end{math}. From our experiments, we found that \begin{math} Threshold =\frac{Max-Min}{2}\end{math} produced the lowest FRR and FAR.

\subsubsection{Comparison of Different Threshold Schemes}
\begin{figure}
\centering
\includegraphics[width=.7\columnwidth]{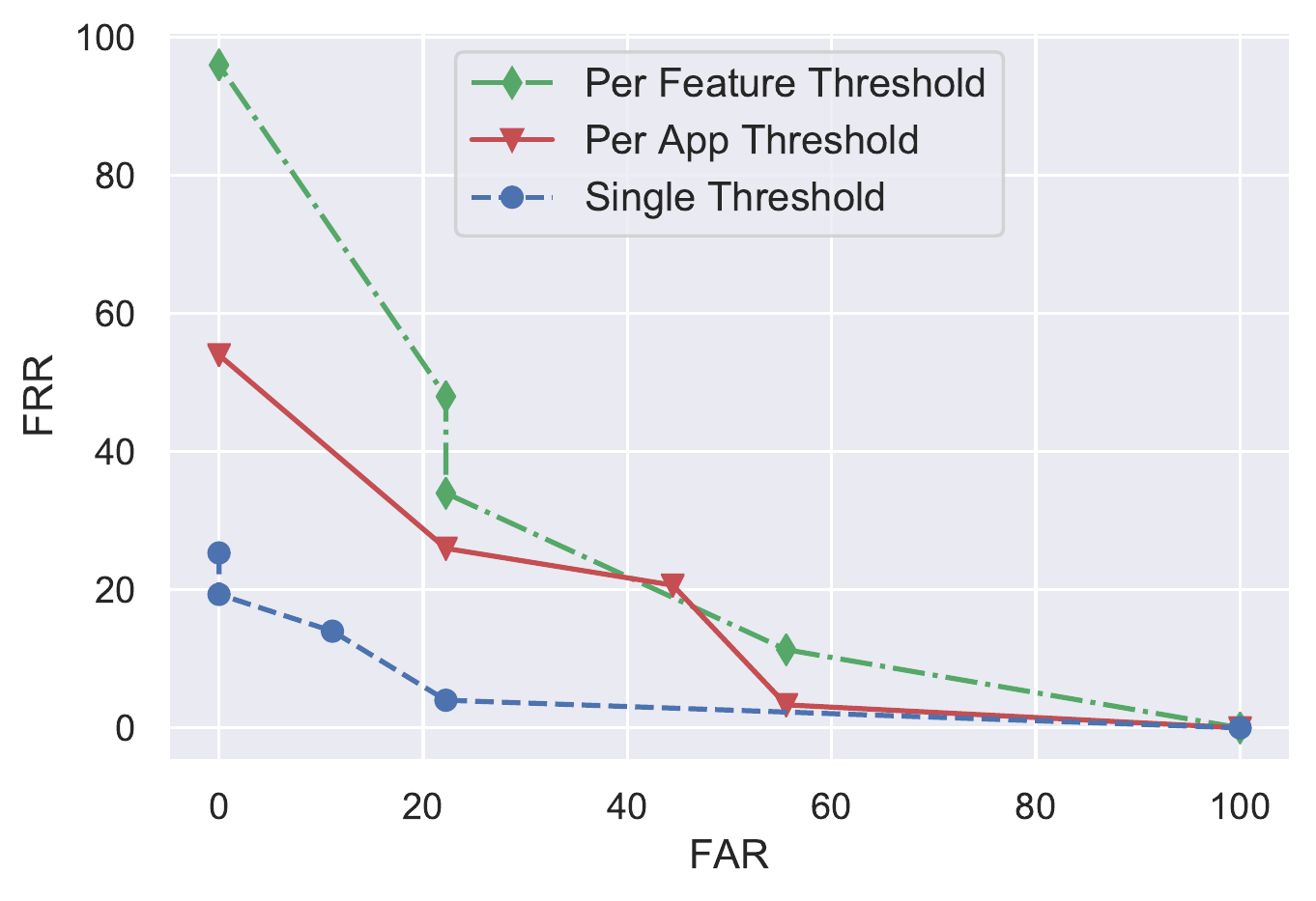}
\caption{FAR vs FRR curve as a function of Polynomial Degree and different threshold approach.}
\label{figure:threshold_compare}
\vspace{-5pt}
\end{figure}

Once we found the optimal threshold for all three threshold schemes, we focused on finding the most efficient threshold technique for our application environment. To compare between different threshold techniques, we picked the optimal threshold for one technique and changed the degree of polynomial from degree 8 to 48 (8, 24, 32, 40, 48) and record the FAR and FRR for these system configurations. For example, for Application Specific Threshold approach, we used $Threshold = \frac{Max-Min}{3}$ and measured FAR and FRR for the five different degree of polynomials. We conducted the same experiments with all the three threshold techniques. \autoref{figure:threshold_compare} illustrates the result of our experiment where we plot FAR against FRR. As our target is to get the lowest FRR and FAR as possible, the line which has the smallest area under it would be the most efficient. Though there were a few applications where FRR was lower for one threshold per feature of each application scheme compared to the other two schemes, from \autoref{figure:threshold_compare}, we can state that in general, the single threshold for all applications approach produces better results compared to the other two techniques.

Based on these findings, we conducted the rest of the experiments for this paper using one single threshold approach with threshold value 57.5 and 32 degree polynomial unless it is mentioned otherwise.



\subsection{Application Behaviors vs Biometrics}
\label{sec:challenges}

The core similarity that makes both application behaviors and biometrics suitable to use with fuzzy extractor is that they both exhibit variability in the data they capture and this data (app behavior or biometric) can be used to uniquely identify an individual (one app instance or one person). The similarities end here and there are several key distinctions that we identified during design and implementation of \mm\ . We discuss the key challenges below.

\vspace{0.25\baselineskip} \noindent\textbf{Noise profiles.} 
The variability in biometric data comes from the inherent errors in the measurement sensor and from environmental factors. For example, fingerprint matching has to account for the fingerprint-sensor noise and for the rotations and translations that happen in practice when a finger is scanned. These types of noises can be modeled as statistical processes and can be accounted in the fuzzy extractor. Applications, by contrast, have variability in behavior over time within a single execution and across executions. Modeling this variability as simple noise is unrealistic and would lead to significant inaccuracies (both false positives and false negatives would be possible in the authentication task). Breaking the behavior modeling process into normalization, encoding, distance metric, and thresholding allows us to adjust to the application variability over time and over executions.

\vspace{0.25\baselineskip} \noindent\textbf{Time-series data.}
Biometrics are typically point-in-time measurements of some physical characteristic. Application behaviors are best represented as time-series data, with both time-varying values and high volume. This implies that using raw sensor data (e.g., log files, firewall statistics) is insufficient as we need to capture instantaneous values and their historical trends.

\vspace{0.25\baselineskip} \noindent\textbf{Normal range of behaviors.}
Application behavior has notoriously varied characteristics, even within an application class like web servers and even for two instances of the same application, deployed in different settings. Biometrics do not change over time and this makes them particularly suitable for authentication. Additional steps are required to project the behaviors of applications into a common range over which we can apply the fuzzy extractor.

To address these challenges, \mm\ adopts several feature transformations suitable for the application behavior domain. First, to normalize the feature values, we use two different normalization techniques. In the first approach, we normalize values across the feature vector using L1 normalization~\cite{l1norm}. In the second approach, we collect two months of data and used MinMaxScaler normalization~\cite{minmax} to assign the normalized value for each feature based on the values it took in two months. The outcome of the normalization is a feature vector with features in the range $[0, 1]$.

Second, to encode the (real-valued) normalized features into a discrete domain, we create an encoding scheme built on the face fuzzy extractor binary mapping proposed by Wang Plataniotis~\cite{wang2007fuzzy}. This encoding converts a $n$-dimensional feature vector to a $n$-dimensional 16-bit binary feature vector. Two random matrices are selected per each application and the distance between the normalized feature vector and each column of the two matrices are quantized into 256 steps representing $x$ coordinates of the feature points. 

The detailed construction is as follows: (1) two random orthonormal matrices $Q_1$ and $Q_2$ of size $n$ $\times$ $n$ are generated; (2) two random vectors $r_1$ and $r_2$ of length $n$ are selected; (3) two matrices $R_1$ and $R_2$ are created by multiplying each column of $Q_1$ and $Q_2$, by each elements of $r_1$ and $r_2$. $R_1$ and $R_2$ are specific to each application and stored in the vault. (4) Vectors $d_1$ and $d_2$ are computed as the Euclidean distance between the application feature vector and each column of $R_1$ and $R_2$. ((5) Each element of $d_1$ and $d_2$ are quantized into 256 steps, generating two vectors $b_1$ and $b_2$. 6) The encoded feature vector $x$ is generated by concatenating $b_1$ and $b_2$, and is stored on the vault as the genuine point $(x, P(x))$ on the polynomial.



\section{System Design and Analysis}
\label{sec:architecture}
In this section, we discuss how  an organization can deploy the \mm\ system in a secure manner, analyze the security of this proposal, and evaluate its performance.

Beyond the question of feasibility (discussed in \autoref{sec:study}), the use of application behaviors for authentication poses a new challenge in terms of security model. While biometrics are useful for authentication because they are considered secret information held by a person and are not too easily copied or replicated\footnote{There are ways to replicate biometrics, but newer techniques for liveness detection (as an example) try to combat such attacks.}, applications are easy to copy, easy to replicate in a new environment, and thus a large part of their behaviors cannot be considered as secret. A key requirement for our architecture is handling the threat of attackers having all the information to mimic application behavior. This restricts the \mm\ approach to application \textit{verification} instead of application \textit{identification}. 

\subsection{Architecture}

The architecture of BAAFE builds around three concepts: a compute location, an application ID, and the application behavior. The goal is to enforce that authentication requests are accepted only from an applications behaving correctly \emph{and} running at the correct compute location.

A compute location is a secure identifier for an environment that can host an application. Compute locations could be physical machines, virtual machines, containers, processes, etc., depending on computing, orchestration, and monitoring infrastructure used. We assume that an approved list of compute locations and expected applications IDs is maintained and secured properly. We require that compute locations must not be forged, meaning that if an attacker compromises an application and injects their own code, he or she cannot pretend to run that attack code from a compute location other than the one the compromised application was running from. This level of security assumes that the host and networking management infrastructure is secure. The purpose of compute locations is to identify the source of behavioral observations reported by sensors.

An application ID is a unique, public identification code assigned to an application, analogous to a username. It could be a static value assigned by the network to each application. The purpose of application IDs is to allow applications to associate an identity with their authentication requests. Finally the application behavior is the sum--total of all observable application activities at a given compute location. Sensors monitor compute locations (e.g., IDSes monitor networked virtual machines) and produce a stream of events that constitutes the application behavior from that location. The purpose of application behaviors is to allow the fuzzy extractor to validate the operation of the application making the authentication request, before releasing the authentication key to the authentication server.

In our threat model, we consider the authentication server, the fuzzy extractor, the sensors, the compute location data, and the communication channels between them to be secure. The trust can be achieved by placing the components within a secure network perimeter, by using mutually authenticated TLS, and by integrating Trusted Platform Module within this components. We note that it is generally easier to secure these components, which have well specified functionalities, that to secure the large number of client applications. 
Sensor security means both that they correctly observe all application activity at a given compute location and that they associate such activity with the correct compute location.




The fuzzy vault, which is a combination of \(n\) genuine points and \(c\) chaff points is not required to be stored in a secured storage as it is resistant to offline brute-force attack. For simplicity we showed in \autoref{figure:baafe-auth} that vault data is provided by the on every authentication request, but this can be optimized by having the fuzzy extractor retrieve the vault data from an application on demand and caching it as needed. Such an optimization does not change the security guarantees, as from the application point of view the vault data is indistinguishable from arbitrary data.

Our architecture offers a choice of behavioral monitors, as it does not impose any particular model. In contrast to our feasibility study where we specified a list of features and their use for the purposes of behavior classification, we expect BAAFE users to have a preferred behavioral monitor, for example in the form of off-the-shelf intrusion-detection systems (IDSes) fit for their applications. To bridge the gap between a given behavioral monitor and the fuzzy extractor, we use explainable AI techniques. In particular, integrated gradients~\cite{10.5555/3305890.3306024} provide a way to derive feature attribution data, explaining the output of the behavioral monitor by a subset of features that contribute the most to that output. The fuzzy extractor then focuses on the values of this concise subset of features, decoupling it from the operation of the behavioral monitor, which could use thousands of features.



\subsection{Analysis}
\label{sec:security-analysis}

The attacker has three capabilities based on the threat model. In increasing order of complexity and strength, an attack can consist of (1) collecting and replaying or synthesizing client application behavior, (2) stealing vault data, (3) compromising the client application.

\vspace{0.25\baselineskip} \noindent \textbf{Collecting and replaying client application behavior.}
The attacker could gain access to one or more client-application behaviors $B_1, \dots, B_t$, by compromising the client host or its network and collecting observations, or by running a similar instance of the client application in their own controlled environment. This attack scenario is fundamentally why application behavior is not similar to biometrics, since biometrics are unique per individual, while application behaviors are replicable to the degree that application code and data can be copied into a new environment. Even with this information, the attacker cannot breach the authentication system. The attacker could try to feed one or more collected behaviors to the fuzzy extractor service, but such an attempt will be rejected since the behavior is not produced by the secure sensors. Additionally the attacker may want to use the application behavior to derive the authentication key $k$ from the fuzzy vault, but without any extra information of the encoder in the fuzzy extractor, this is equivalent to a brute-force attack on the key space. 

\vspace{0.25\baselineskip} \noindent \textbf{Stealing of vault data for offline analysis.}
The attacker could obtain the vault data $V$ (but not control of the fuzzy extractor service itself) to perform an offline attack. In this scenario the attacker may attempt to reconstruct possible keys, but since the fuzzy vault does not contain information to validate the candidate keys, the attacker has to try them against the authentication server in an online fashion. The attacker could derive key candidates by randomly selecting $d+1$ points  from the vault, interpolating a polynomial, reconstructing the key, and validating the key through an authentication attempt. For a polynomial of degree $d$ and a vault of size $n+c$ (total genuine and chaff points), the attacker will need to make
$\frac{1}{2}{n+c \choose d+1}$
attempts in expectation, to reveal the secret. By choosing the number of chaff points to be large enough, we can achieve any desired level of security and thus make such reconstruction attacks infeasible.

The above expectation value holds for the case of global threshold and per-application threshold. In the case of per-feature thresholds, the vault does not simply contain $n+c$ points but rather it is structured into $m$ subsets for $m$ number of features, each with $n_i+c+i$ points for $1\leq i\leq m$. Even though the total number of points remains the same, $n+c=\sum_{1\leq i\leq m}n_i+c_i$, the added structure gives the attacker additional information and the number of expected brute-force attempts lower than before:
$\frac{1}{2}\sum_{m \choose d+1}\prod_{x_{a_1},\dots,x_{a_{d+1}}}(n_{x} + c_{x})$.
Increasing the number of chaff points similarly allows us to achieve the desired level of security.  



\vspace{0.25\baselineskip} \noindent \textbf{Compromising the client application.}
The attacker could gain control of the client application and its environment on the client host. If the attacker replaces the client application with their own code or other modifies the client application, the behavior will deviate from normal, features will fall outside of the expected threshold, and any authentication attempts will fail. The only option the attacker has at this point is not to deviate from normal application behavior or to replay an observed or known application behavior through actions on the compromised host. In such a case, we actually achieve the goal of authenticating an application in a given compute location only if its behavior is normal, even though the application is nominally controlled by the attacker. We note that this security guarantee is relative to the accuracy of the behavioral model used by the monitor.

\begin{figure}
	\centering
	\subfloat[Effect of chaff-point count on enrollment time.\label{figure:enroll_chaff_time}] {
		\includegraphics[width=0.67\columnwidth]{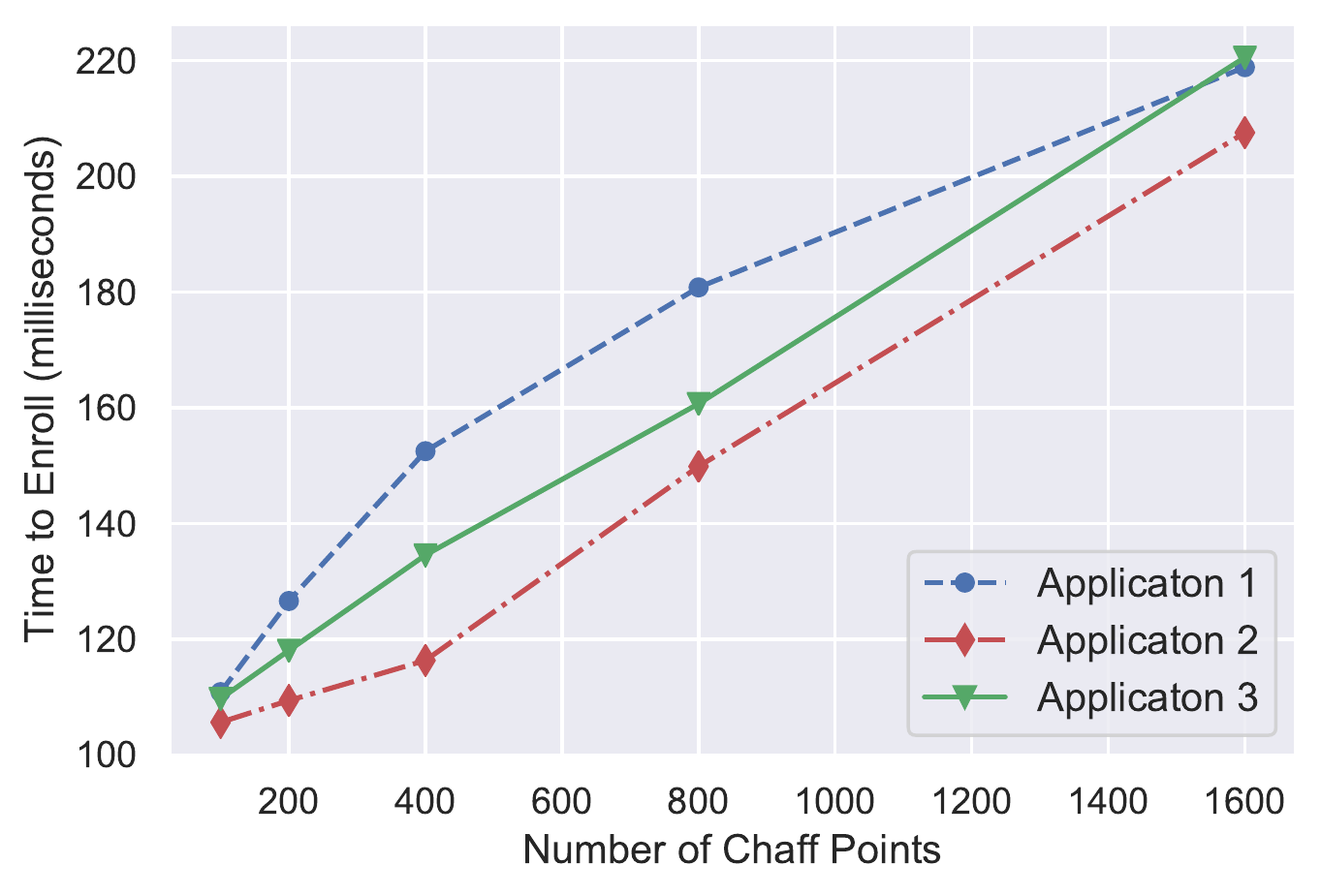}
	}

	\subfloat[Effect of polynomial degree on enrollment time.\label{figure:enroll_degree_time}] {
		\includegraphics[width=0.67\columnwidth]{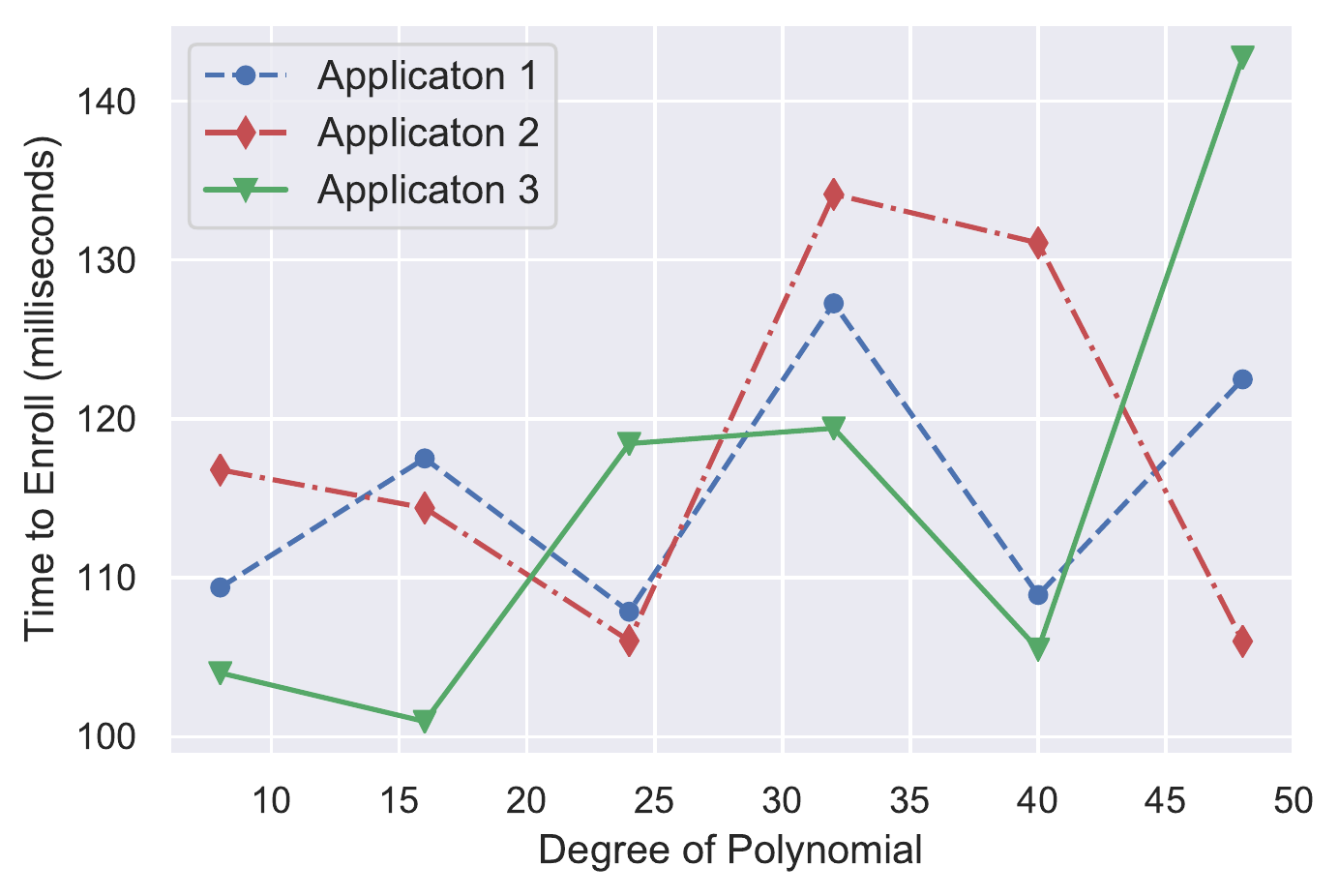}
	}

	\caption{Latency of enrollment is affected by the number of chaff points but not the degree of the polynomial.}
\end{figure}

\subsection{Performance Evaluation}

The performance of enrollment and authentication schemes were measured on a Dell
Precision laptop, running Windows 10, with an Intel Core I7 2.60 GHz CPU and 32GB of RAM. All of the schemes were developed in Python 3 and the Sage math library~\cite{sagemath} for finite field operations. 

\begin{figure*}
	\centering
	\subfloat[Effect of chaff point count on successful authentication time.\label{figure:auth_success_chaff_time}] {
		\includegraphics[width=.31\textwidth]{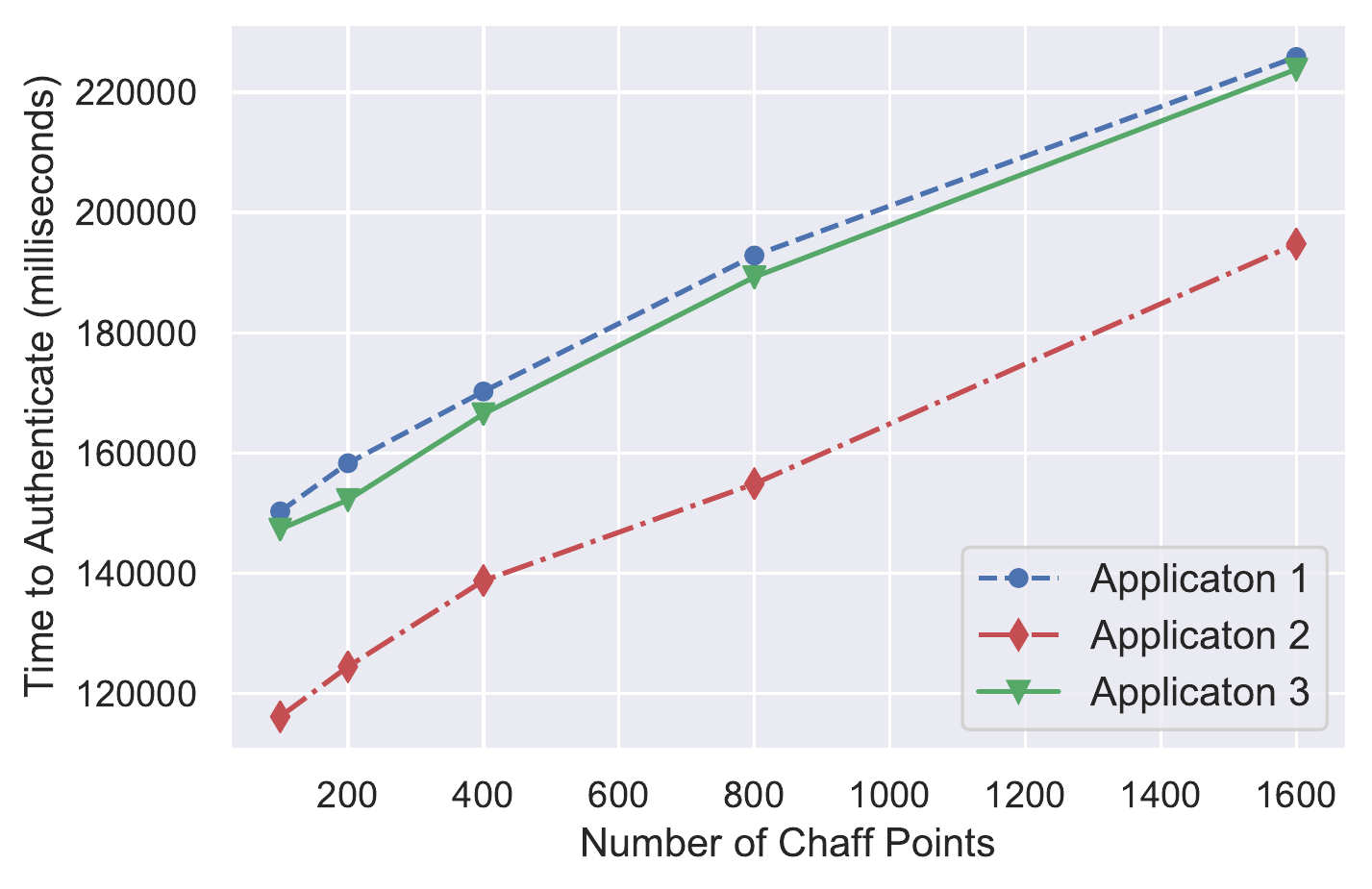}
	}
	\hfill
	\subfloat[Effect of polynomial degree on successful authentication time.\label{figure:auth_success_degree_time}] {
		\includegraphics[width=.30\textwidth]{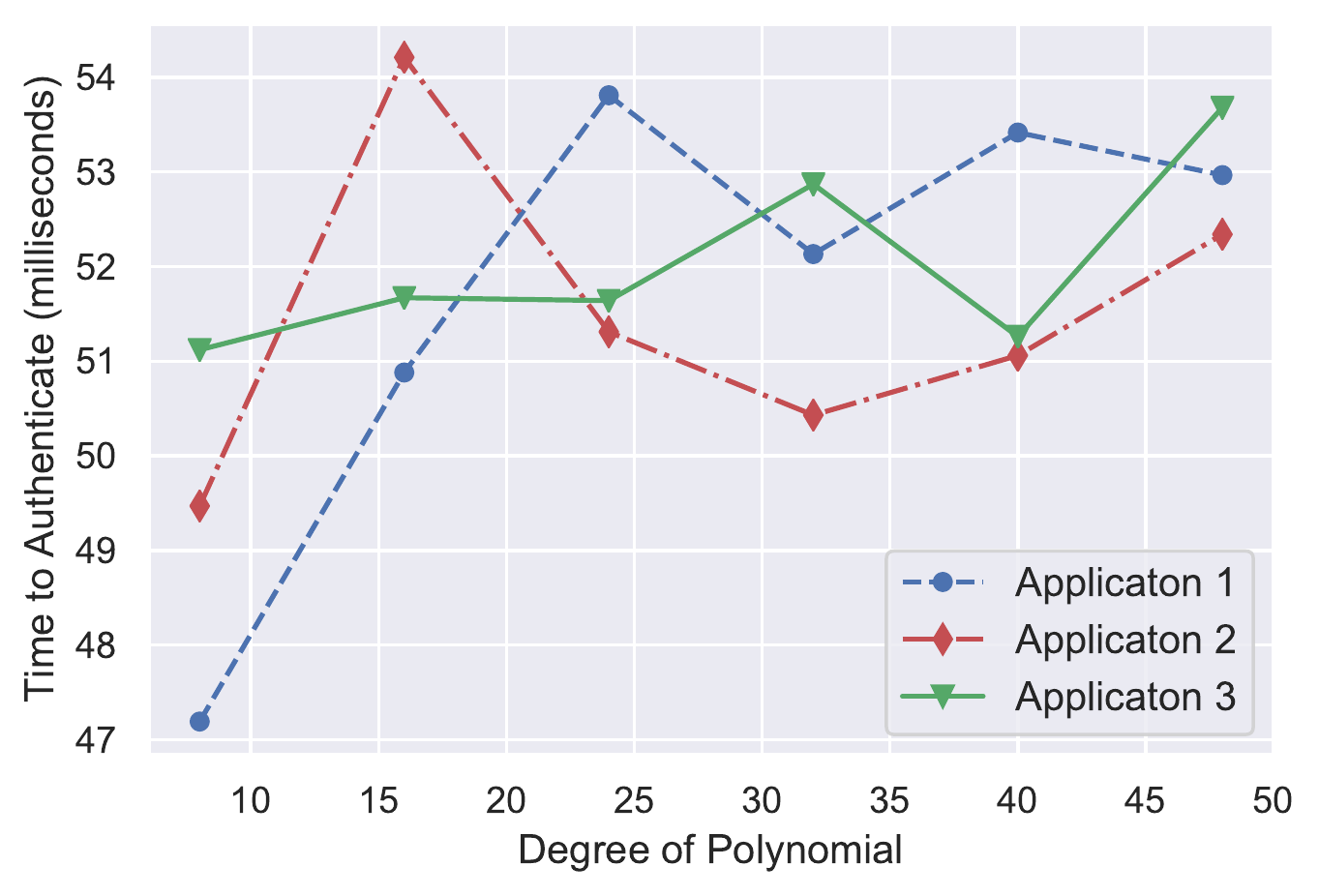}
	}
	\hfill
	\subfloat[Effect of polynomial degree on failed authentication time.\label{figure:auth_failed_degree_time}] {
		\includegraphics[width=.31\textwidth]{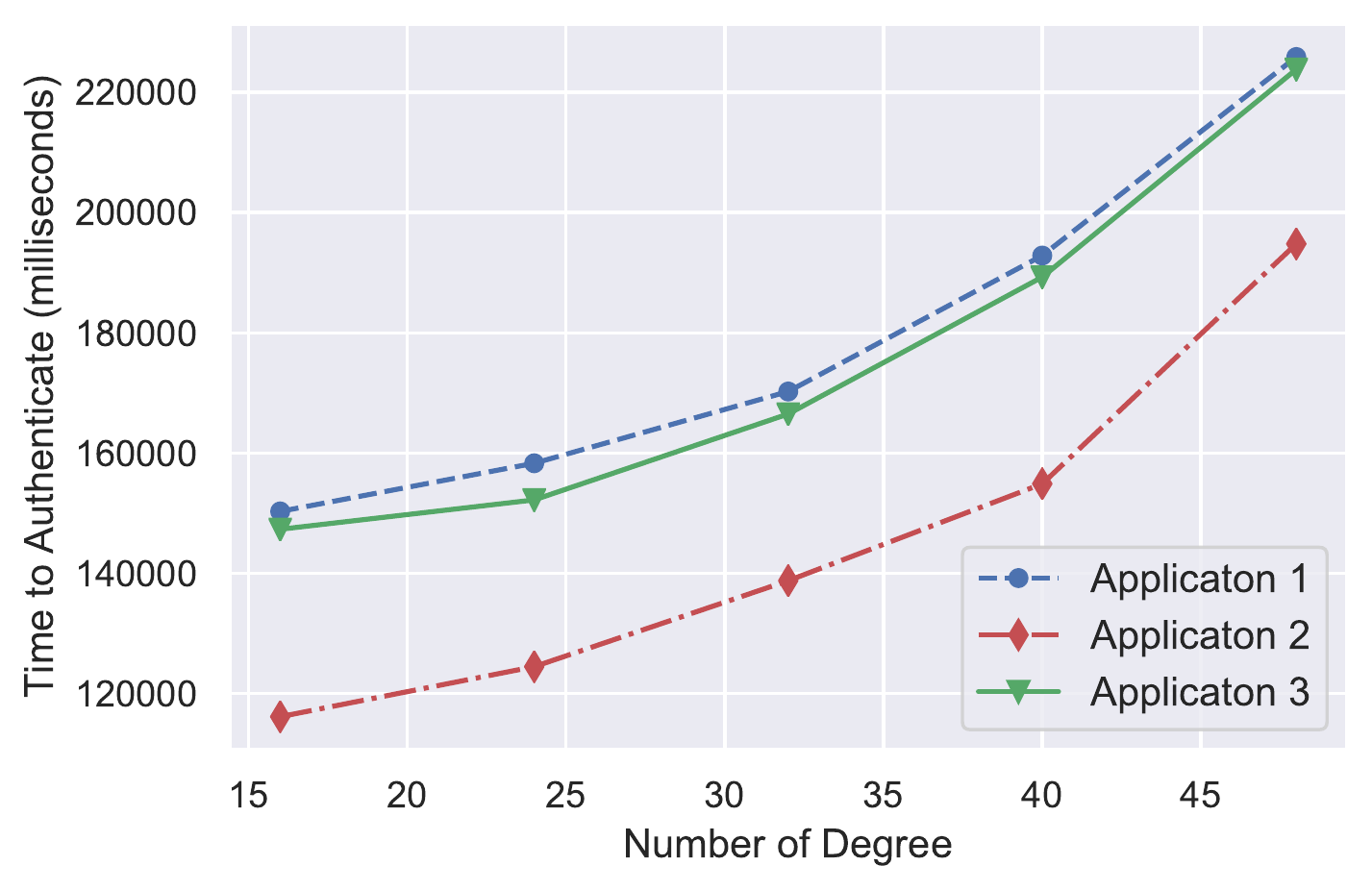}
	}
	\caption{Latency of authentication is impacted by number of chaff points in use, but not by the degree of the polynomial.}
\end{figure*}

\vspace{0.25\baselineskip} \noindent \textbf{Enrollment.}
For the  performance evaluation of the enrollment process, we identified the effects of number of chaff points and degree of polynomial on the performance of enrollment. We randomly picked three applications from the set of 10 applications for this experiment as the type of an application does not affect the enrollment process.  For each application, we executed the enrollment process five times and calculated the average enrollment time. 

\emph{Effect of number of chaff points on enrollment.}
We measured the average enrollment time for the three randomly picked applications for varying number of chaff points (100, 200, 400, 800, and 1600). 
\autoref{figure:enroll_chaff_time} shows that for each application, enrollment time gradually increased with the increase in number of chaff points. This is expected as the process of adding chaff points to a fuzzy vault has time complexity linear in number of chaff points.


\emph{Effect of degree of polynomial.} For this experiment, we used 200 chaff points and measured the average enrollment time for the three randomly picked applications by changing the degree of polynomial from 8 to 48 (8,16,24,32,40,48). As shown in \autoref{figure:enroll_degree_time}, there is no direct relation between degree of polynomial and enrollment time.

\vspace{0.25\baselineskip} \noindent \textbf{Authentication.}
Performance of the authentication process is particularly important since this  has a direct implication on the security from the brute force attack. In this experiment, we identified the effects of number of chaff points and degree of polynomial on the successful and unsuccessful authentication attempts. First, we randomly picked three applications from the set of 10 applications. To gauge the performance of successful attempts, we measured the authentication time by executing the authentication process five times for each application against the related fuzzy vault/helper data of the application and calculated the average time of successful authentication attempt. We executed the authentication process five times of one application against the fuzzy vault of a completely different application and calculated the average time of authentication for the unsuccessful attempts.

\textit{Effect of number of chaff points.} 
\autoref{figure:auth_success_chaff_time} presents that the successful authentication time increased with the increase of number of chaff points. During authentication, the algorithm described in \autoref{fig:recon} searches for nearest points for each input feature, where the search space increases with the increase in the number of chaff points. Hence, the overall runtime of algorithm has linear time complexity in terms of number of chaff points which is depicted in \autoref{figure:auth_success_chaff_time}. We noticed the same behavior for unsuccessful authentication as well.

\emph{Effect of degree of polynomial.}
For the \textit{successful authentication}, we did not find any relation between the degree of polynomial and the authentication time as depicted in \autoref{figure:auth_success_degree_time}. From  \autoref{fig:recon}, we noticed that after finding the closest points, randomly selecting $d+1$ points to reconstruct the polynomial of degre $d$ caused an unpredictable number of trials to authenticate an application and varying authentication time. Even for the same degree of polynomial, the number of trial changed and produced varying authentication times.

For the \textit{unsuccessful authentication}, we set the number of trial limit as twenty thousand and tried to authenticate one application using the fuzzy vault of another application. As illustrated in \autoref{figure:auth_failed_degree_time}, we see a clear relationship between the completion of an unsuccessful authentication attempt and the degree of polynomial. The higher the degree of a polynomial, the more time it took to complete the authentication process for a given input value. 

From this experiment we can also estimate the time to execute a successful brute force attack on our proposed system. With a degree-32 polynomial and 200 chaff points, it may take \begin{math}
1.51 \times 10^{73}
\end{math} years to exhaust all the combinations of points using a computer similar to what is used in this study.

\vspace{0.25\baselineskip} \noindent \textbf{Authentication Accuracy.}
We investigated FAR and FRR when varying the degree of the polynomial.
 
\begin{figure}
	\centering

	\subfloat[Effect of polynomial degree on FAR.\label{figure:far}] {
		\includegraphics[width=0.7\columnwidth]{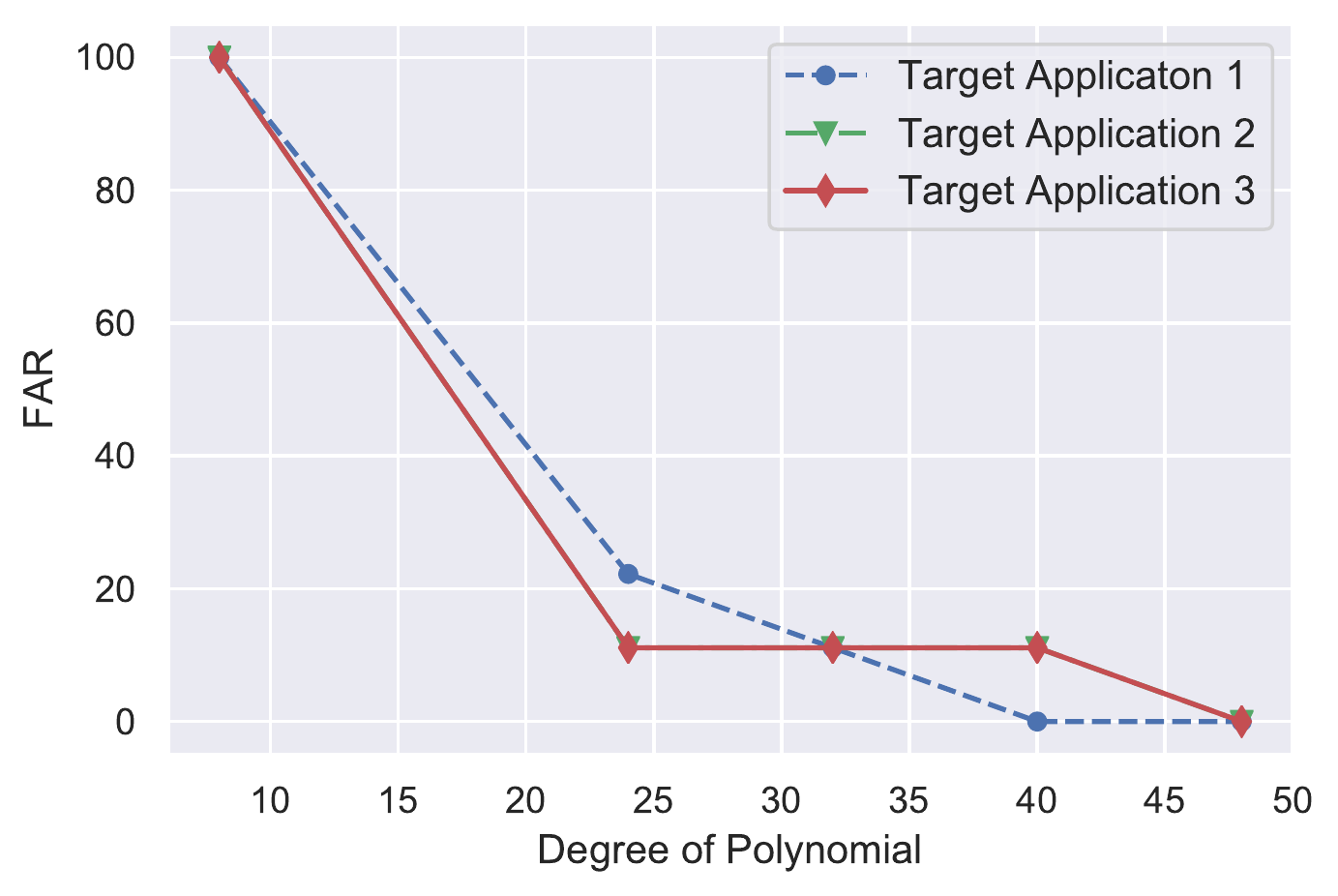}
	}

	\subfloat[Effect of polynomial degree on FRR.\label{figure:frr}] {
		\includegraphics[width=0.67\columnwidth]{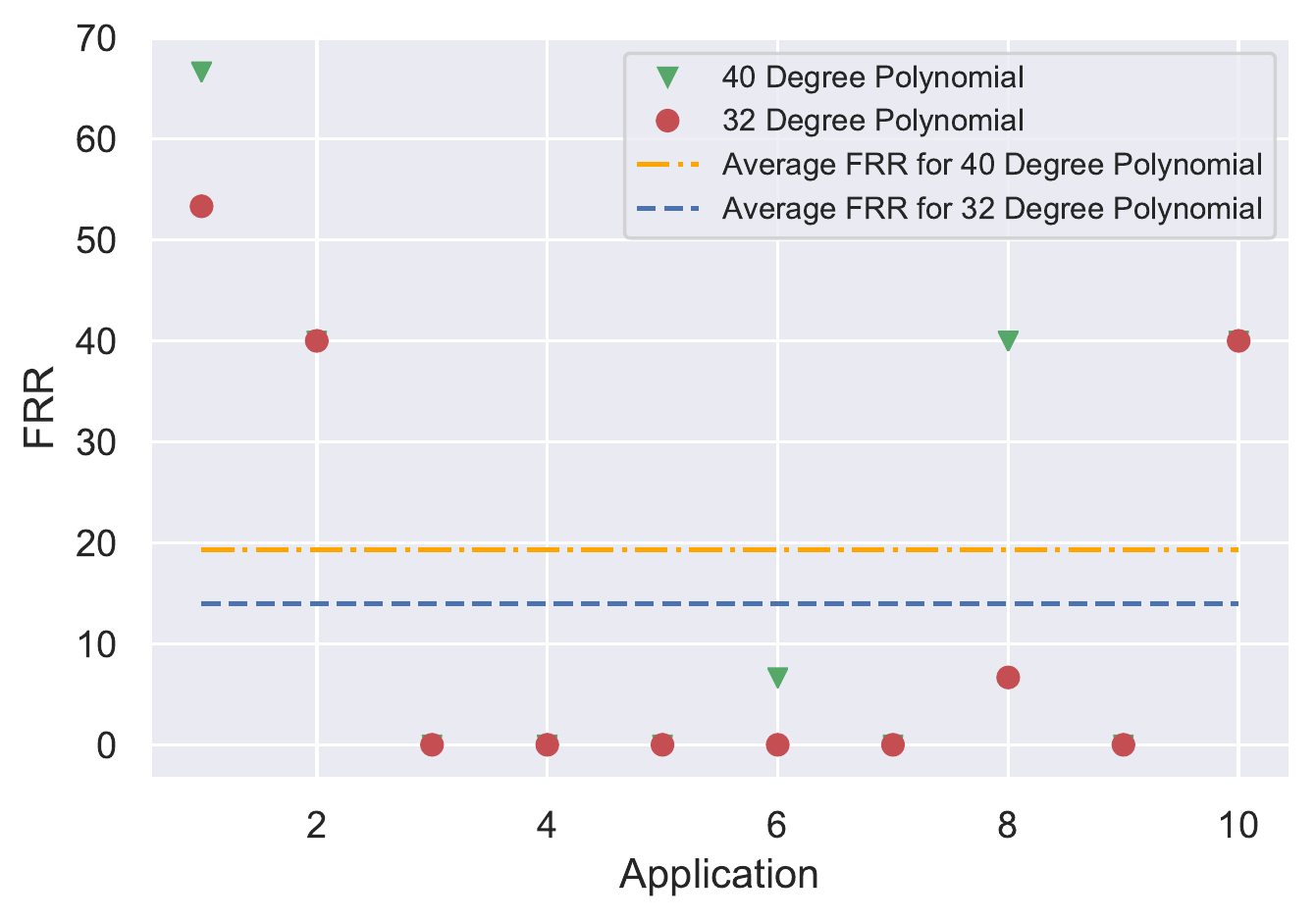}
	}

	\caption{Accuracy is influenced by the degree of the polynomial.}
\end{figure}

We repeated the FAR calculation process  for different degrees of polynomials (8, 24,32, 40, 48). \autoref{figure:far} shows the results of this experiment. We noticed that FAR was 100\% with 8 degree of polynomial and started to reduce as we increased the degree of polynomial. 

100\% FAR with degree-8 polynomials means all the 9 applications have at least 9 features, which are close to the features of Application 1. When we increase the degree of polynomials, we are effectively considering more data points, which  reduced the FAR. We noticed 0\% FAR with degree-40 polynomial when the target application is Application 1.

We repeated the same process for Applications 2 and 3. Both of these applications have another deployment in our dataset of 10 applications but are serving the same services to the clients. With these two target applications, we noticed that FAR was 11.11\% starting from degree-24 polynomials and the only application that successfully authenticated as the target application is the clone of the respective application, which was running on a different environment. We noticed 0\% FAR with degree-48 polynomials. These results explain that if an application has multiple deployments, we may need higher-degree polynomials to distinguish them, as many of the data points between the two deployments can be similar even if the applications are running on two different environments.

\autoref{figure:frr} represents FRR for each application for different degrees of polynomials. 
We noticed that with higher-degree polynomials, FRR became higher for some applications. For five applications, FRR was found 0\% for degree-32 and degree-40 polynomial. For  Application 8, FRR was reduced from 40\% to 6.67\% when we changed the degree of polynomial from 40 to 32, and for Application 6, FRR was reduced from 6.67\%  to 0\% with the change in degree of polynomial. 

For two of the applications, where FRR remained high even with a lower-degree polynomial, we picked one such application and increased the enrollment frequency before executing the authentication process. We found that if we re-enrolled every five days, FRR becomes 0\% for that application. This behavior explains that some of the applications' behavior were not stable over a month and needed re-enrollment to reduce the FRR.


\section{Discussion and Future Work}
\label{sec:discussion}


\smallskip \noindent \textbf{Behavioral classification v. behavioral authentication.} The focus of our work was construction of authentication keys using application behavior whereas in behavioral classification the main focus is to classify applications as the name suggests. Our motivation in building this system is to draw a line between the authentication server and the proposed system. Such distinction helps to avoid unwanted changes on servers that are already secure. Besides, in case the client application and the service belong to two different entities, the server does not learn about the behavior of the client. 

\smallskip \noindent \textbf{Future studies.}
Our current evaluation targets 10~applications running in our organization's network. A few of these applications had similar traits, for example some of the applications in our data set serve web services and two applications are antivirus solution for different environments. For these 10~applications we collected data for the duration of 2 months from our data lake and security incident and event management (SIEM) system.
In light of the promising results on this data, we are motivated to expand the study to 1) cover extra features to achieve higher accuracy, 2) run the experiment for a longer duration of time to study changes in the application behavior over time, and 3) run the experiment for a larger set of applications with similar behavior. Based on our current experiment, we found that when we went from a single threshold for all application, to application-specific and later feature-specific thresholds the authentication accuracy decreased. We also noticed that although the feature-specific threshold scheme generated less accuracy on average, for some applications the feature-specific threshold produced better FRR than the single-threshold scheme. In the future, we want to explore more the threshold schemes and design the \mm\ system to adapt its threshold scheme based on workload.

\smallskip \noindent \textbf{Integration into Existing Applications.} In some application domains, such as database management systems, the clients performs an initial authentication to create an authenticated connection to the server (or even a pool of authenticated connections). From that point on, the client does not need to re-authenticate as long as it uses one of the already opened, authenticated connections. The motivation for such a design is performance as it is much faster to reuse an authenticated, open connection than to open a new connection and authenticate each time a client wants to connect to the database server. However, this comes with the price of creating new attack vectors from having the idle, pre-authenticated connections available to the attacker. This design is not immediately compatible with \mm, and we intend to extend our system in the future to allow such performance optimization.

\smallskip \noindent \textbf{Security--Performance Tradeoff.}
As shown by the performance evaluation (Figures~\ref{figure:auth_success_chaff_time} and~\ref{figure:auth_failed_degree_time}), the time to authenticate depends on the number of chaff points and the degree of the polynomial used by the fuzzy extractor, both of which have security implications. Further investigation is needed to identify the optimal point of this tradeoff.


\section{Conclusion}
\label{sec:conclude}

We introduced a privacy-preserving application behavioral authentication method using a fuzzy-extractor construction. 
The enrollment protocol secret-shares the authentication key over a $d$-degree polynomial and generates public vault data consisting of true polynomial points and fake off-polynomial ``chaff'' points. The authentication protocol reconstructs the correct secret key only if the provided features match the enrollment features, with some error tolerance. We analyzed security of our technique and show that 1) attacker who compromises the client cannot learn the key, 2) attacker who compromises the vault can learn neither the key nor behavior and 3) attacker who observes feature values cannot mimic client's behavior. Our evaluation of 10 applications demonstrates a promising result of 0\% False Accept Rate with an average False Rejection Rate 14\% and takes about 51 milliseconds to successfully authenticate a client application.


\bibliographystyle{IEEEtranS}
\bibliography{all}


\section{Appendix}
\label{sec:appendix}

Feature list used in the feasibility study of \autoref{sec:study}:

\begin{itemize}[noitemsep, leftmargin=*]
\item The number of unique URLs accessed by an application.
\item The number of unique categories of URLs that are accessed by an application.
\item The total data in bytes received by an application from all sources.
\item The total data in bytes sent by an application to all destinations.
\item The number of HTTP requests initiated by an application.
\item The number of SSL requests initiated by an application.
\item The number of proxy authentication failed by an application.
\item The number of responses received with HTTP Code 200.
\item The number of Firewall requests allowed for an application.
\item The number of Firewall requests denied for an application
\item The number of  outgoing connections from the application hosting machine.
\item The number of unique destinations based on all the outgoing connections from the application hosting machine.
\item The number of unique source ports based on all the outgoing connections from the application hosting machine.
\item The number of requests that an application made to the enterprise SIEM.
\end{itemize}

\begin{figure}[h!]
	\centering
	\subfloat[Successful authentication attempt \label{figure:vault_success_viz}] {
		\includegraphics[width=0.7\columnwidth]{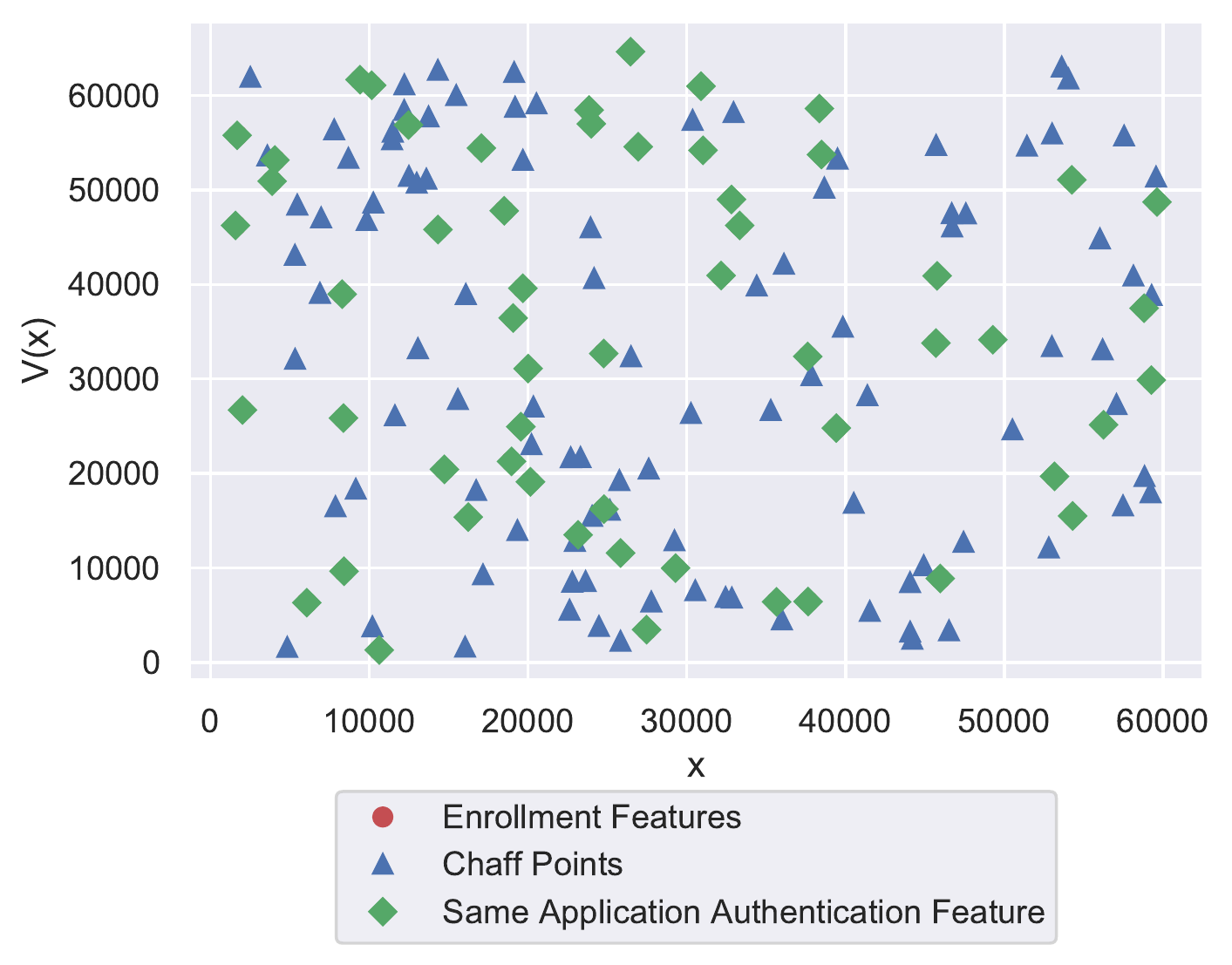}
	}

	\vspace{0.5\baselineskip}
	\subfloat[Failed authentication attempt \label{figure:vault_failed_viz}] {
		\includegraphics[width=0.7\columnwidth]{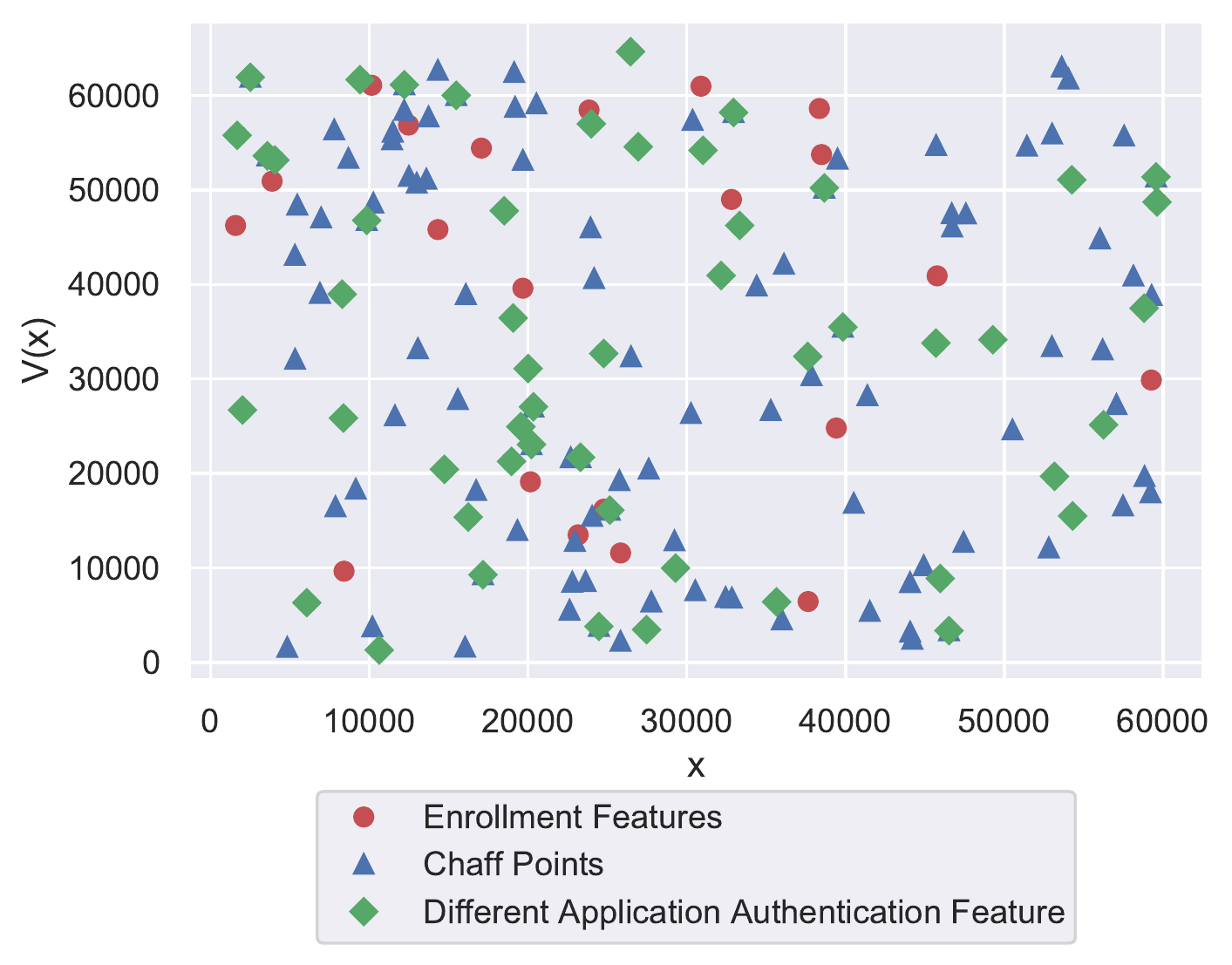}
	}

	\caption{Visualization of vault point matching during key reconstruction.}
\end{figure}

\end{document}